\documentclass[]{spie}  
\usepackage[]{graphicx}

\title{Precision near-infrared radial velocity instrumentation II: Non-Circular Core Fiber Scrambler} 

\author{Peter P. Plavchan\supit{a}, Bottom, M.\supit{b}; Gao, P.\supit{b}; Wallace, J. K.\supit{c}; Mennesson, B.\supit{c}; Ciardi, D.\supit{a}; Crawford, S.\supit{c}; Lin, S.\supit{c}; Beichman, C.\supit{a}; Brinkworth, C.\supit{a}; Johnson, J.\supit{d};  Davison, C.\supit{e}; White, R.\supit{e}; Anglada-Escude, G.\supit{f}; von Braun, K.\supit{g}; Vasisht, G.\supit{c}; Prato, L.\supit{h}; Kane, S.\supit{i}; Tanner, A.\supit{j}; Walp, B.\supit{k};  Mills, S.\supit{l};    
\skiplinehalf
\supit{a}NASA Exoplanet Science Institute, California Institute of Technology, 770 S Wilson Ave, Pasadena, CA, USA; \\
\supit{b}California Institute of Technology\\
\supit{c}Jet Propulsion Laboratory\\
\supit{d}Harvard University\\
\supit{e}Georgia State University\\
\supit{f}University of Goettingen\\
\supit{g}Max Planck Institut Astronomie, Heidelberg\\
\supit{h}Lowell Observatory\\
\supit{i}San Francisco State University\\
\supit{j}Mississippi State University\\
\supit{k}SOFIA \\
\supit{l}University of Chicago
}

\authorinfo{Further author information: (Send correspondence to P.P.P.)\\P.P.P.: E-mail: plavchan@ipac.caltech.edu, Telephone: 1 626 234 1628}

\begin{document} 
  \maketitle 

\begin{abstract}

We have built and commissioned a prototype agitated non-circular core fiber scrambler for precision spectroscopic radial velocity measurements in the near-infrared H band. We have collected the first on-sky performance and modal noise tests of these novel fibers in the near-infrared at H and K bands using the CSHELL spectrograph at the NASA InfraRed Telescope Facility (IRTF). We discuss the design behind our novel reverse injection of a red laser for co-alignment of star-light with the fiber tip via a corner cube and visible camera.  We summarize the practical details involved in the construction of the fiber scrambler, and the mechanical agitation of the fiber at the telescope.  We present radial velocity measurements of a bright standard star taken with and without the fiber scrambler to quantify the relative improvement in the obtainable blaze function stability, the line spread function stability, and the resulting radial velocity precision.  We assess the feasibility of applying this illumination stabilization technique to the next generation of near-infrared spectrographs such as iSHELL on IRTF and an upgraded NIRSPEC at Keck.  Our results may also be applied in the visible for smaller core diameter fibers where fiber modal noise is a significant factor, such as behind an adaptive optics system or on a small $<$ 1 meter class telescope such as is being pursued by the MINERVA and LCOGT collaborations.

\end{abstract}


\keywords{exoplanets, instrumentation, near-infrared spectroscopy, radial velocity surveys}

\section{INTRODUCTION}
\label{sec:intro}  

Visible Doppler spectroscopy is the most successful technique used to detect or confirm the majority of $\sim$900 known exoplanets (Fig 1)\cite{akeson13}.  The 2012 NSF Astronomy Portfolio Review made the detection of Earth-mass planets in habitable zone orbits with the radial velocity method the \#1  science priority for Planetary Systems and Star Formation (PSSF):  

\begin{quotation}
NWNH [New Worlds New Horizons 2010 Astronomy Decadal Survey] recommends the development of new spectrographs capable of achieving 0.1-0.2 m/s precision, and adequate allocation of observing time on 4-m to 10-m telescopes. \dots Critically needed technical capabilities [are] instrument development for extreme-precision optical spectrographs (to reach 10 cm/s) and high-resolution near-IR spectrographs to detect planets orbiting cool stars \dots  Coupling of light to the instrument, opto-mechanical stability and optimal wavelength calibration are all areas that still merit work. Thus, substantial instrument and analysis development will likely be needed, over the course of several years.
\end{quotation}

\begin{figure}[tb]
  \begin{center}
    \includegraphics[width=0.80\textwidth]{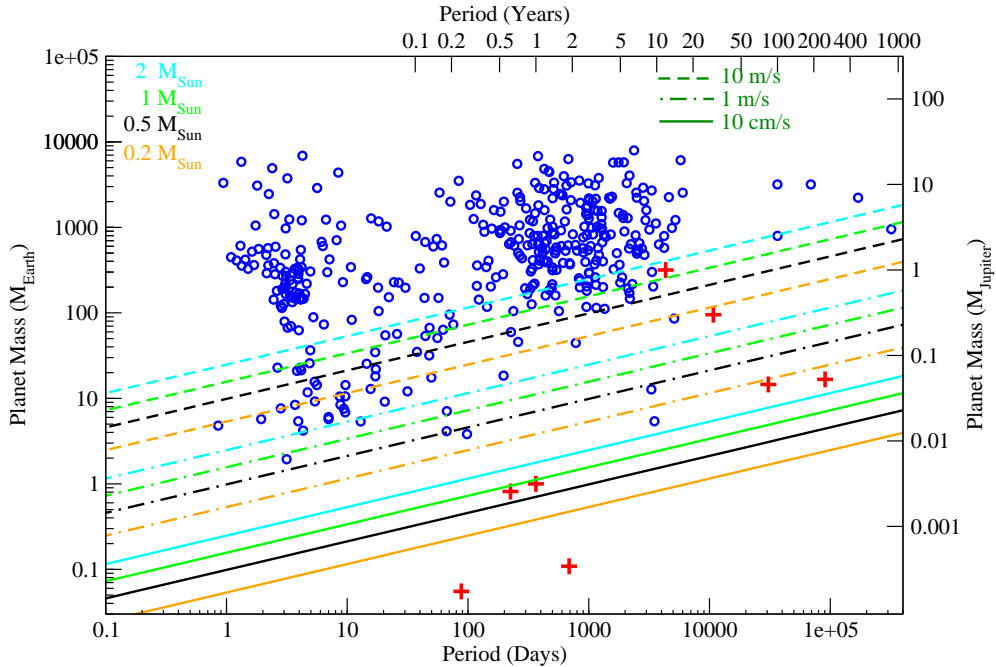}    
      \end{center}
      
  \caption{Confirmed exoplanets shown as blue circles from the NASA Exoplanet Archive \cite{akeson13} plotted as a function of orbital period on the horizontal axis, and planet mass on the vertical axis. Solar System planets are shown as red crosses.  The dashed, dot-dashed, and solid lines correspond to the sensitivity limits for radial velocity precisions of 10 m/s, 1m/s and 10 cm/s respectively, and the colors of these lines correspond to different stellar host masses of 2 M$_\odot$ (blue),  1 M$_\odot$ (green), 0.5 M$_\odot$ (black), and 0.2 M$_\odot$ (orange).  In the absence of red noise due to stellar activity or systematic noise sources, exoplanets above these lines can be detected for a given stellar mass and radial velocity precision.\label{fig:f1}}
\end{figure}

The current generation of visible precision radial velocity instrumentation achieves $\sim$1 m/s single measurement precision.  The future potential of the precision radial velocity method is exemplified by the recent detection of an Earth-mass exoplanet (radial velocity semi-amplitude K=51 cm/s) with a 3.2-day orbit around $\alpha$ Cen B with the HARPS-South spectrograph \cite{dumusque12}.  Besides astrophysical limitations due to radial velocity jitter from stellar activity, there are two fundamental challenges for precision radial velocity instrumentation.  The first challenge is wavelength calibration, since the Doppler method relies on the detection of subtle wavelength shifts in stellar spectra on the order of 1/1000$^{th}$ of a detector spectral pixel.  Wavelength calibration is addressed with two approaches -- one, a common optical path absorption gas cell that provides a relative wavelength calibration for a classical spectrograph such as HIRES \cite{HIRESref}; and two, Thorium Argon emission lamps, stabilized Fabry-Perot etalons, or laser frequency combs that provide absolute stable wavelength references for a pressure, gravity and temperature stabilized spectrograph such as HARPS-S \cite{HARPSSref,FabryPerotref,osterman10}.  

The second challenge for precision radial velocity instrumentation is stabilizing the input illumination of the spectrograph.  Spectrograph optics such as the echelle grating, collimators, cross-disperser and camera are not ideal in practice and feature inhomogeneities.  Consequently, how optics are illuminated and, more importantly, how that illumination changes with time can introduce undesirable shifts in spectra on the detector and changes to the line-spread function (LSF).  These factors ultimately introduce systematic radial velocity errors.  This challenge is addressed with HIRES by relying on excellent telescope guiding.  For comparison, HARPS-S injects starlight into a fiber double-scrambler consisting of two circular core step-index fibers and a far-field -- near-field exchange before the output illuminates the stabilized spectrograph\cite{hunter}.  The output near- and far-field spatial distribution from circular core fibers can have a strong dependence on the input illumination -- e.g., guiding the star on the center of the input fiber tip can produce a fiber output with an approximately Gaussian far-field spatial distribution of starlight; guiding the star towards the edge of the fiber tip can instead produce a fiber output with an annular far-field spatial distribution of starlight.  

How are we going to address these principle challenges?  Non-circular core fibers are promising for scrambling the starlight to stabilize the spectrograph input illumination, and have been demonstrated to produce output illumination in the near-field that depends minimally on the input illumination\cite{spronck,spronck2,spronck3,spronck4}.  Fibers with square, rectangular, octagonal and more complicated core geometries are a relatively recent advance driven by pump laser efficiency in applied physics research.  These fibers are already finding use on HARPS-N and the HIRES fiber scrambler (Andrew Howard, priv. comm.).  However, even with the spatial mode scrambling provided by the novel geometry of the fibers, manual agitation may still be necessary to mitigate the negative effects of modal noise -- the variation in illumination intensity in different propagating modes in the fiber (${\S}$3.6).  The number of propagating modes in a fiber is a function of the inverse of wavelength and the fiber diameter.  Thus, fiber modal noise is particularly problematic for near-infrared precision radial velocity instrumentation such as the Habitable Zone Planet Finder\cite{2012SPIE.8446E..8JM}, and for planned robotic small telescope radial velocity spectrograph systems such as LCOGT and MINERVA which have a small etendue and thus smaller fiber diameters \cite{bottom13}. 

Radial velocity measurements in the near-infrared are advantageous for many reasons including the potential detection of terrestrial mass planets around cool M dwarfs.  However, radial velocity precision in the near-infrared historically lags the visible \cite{bean10,bailey11}.   Fortunately, a number of facilities are funded to become operational in the next five years, including the Habitable Zone Planet Finder, CARMENES, SPIRou, iGRINS, and iSHELL \cite{HZPFref,CARMENESref,iGRINSref,iSHELLref}.  The Habitable Zone Planet Finder, SPIRou and CARMENES are purposely built for precision radial velocity measurements.  Additionally, recent advances have pushed the achievable radial velocity precision in the near-infrared closer to parity with the visible.  These advances include the use of ammonia and isotopic methane absorption gas cells the achieve a precision of $\sim$5-10 m/s \cite{bean10,anglada12,plavchan13}, and the prototype for the fiber-fed Habitable Zone Planet Finder, Pathfinder, which made use of a novel laser frequency comb and Uranium-Neon lamps for wavelength calibration to achieve a precision of $\sim$15 m/s \cite{osterman10,PATHFINDERref}.   The next generation of near-infrared facilities hold the potential to achieve $\sim$1-3 m/s single measurement precision, which is sufficient to detect terrestrial planets in the habitable zones of M dwarfs.  Active areas of technology development include Uranium-Neon emission lamps, stabilized Fabry-Perot etalons, and single-mode fiber-based spectrographs behind adaptive optics equipped telescopes\cite{redman,FabryPerotref}. 

In this paper, we present the design, construction and first light of a prototype agitated non-circular core fiber scrambler to stabilize the input illumination of the near-infrared high resolution spectrograph CSHELL on the NASA InfraRed Telescope Facility (IRTF) \cite{tokunaga1990,greene1993}.  Given the costs of new high-resolution cryogenic spectrographs, we have taken the low-cost approach of modifying existing near-infrared spectrographs for enabling high precision radial velocity measurements through the use of absorption gas cells and a fiber scrambler.  These instruments are available to the community to use on CSHELL, and interested parties are encouraged to contact the authors for more information.  The first light paper with the isotopic methane absorption gas cell was presented in Anglada-Escude et al. (2012).    The companion SPIE paper by Plavchan et al. (2013) goes into more depth on the scientific motivation for near-infrared observations and the current state of the field, our gas cell instrumentation, as well as updating the current progress of our survey data analysis and techniques.   A future publication, Bottom et al. in prep., will present detailed science results from fiber scrambler prototype.

\section{Spectrograph}

CSHELL is the ``host'' spectrograph for our ``symbiotic'' instrumentation.  The spectrograph is almost 20 years old, with (non-simultaneous) wavelength coverage from 1-5.5 $\mu$m, and with a Hughes SBRC 256x256 InSb CCD\cite{tokunaga1990,greene1993}.  The instrument is mounted at the Cassegrain focus of IRTF on the back of the primary mirror support structure.  CSHELL captures a single $\sim$5 nm order of spectra at 2.3 $\mu$m at a resolution of R$\sim$46,000 with a 0.''5 slit.  For comparison, the non-cross-dispersed CRIRES on the VLT had a spectral grasp of $\sim$50 nm of a single order that spans four CCDs, and has recently added a cross-disperser to increase the spectral grasp further.  Thus, the spectral grasp of CSHELL is limited compared to more modern spectrographs.  Since iSHELL is a funded replacement for CSHELL, we are graciously allowed by Dr. John Rayner and Dr. Alan Tokunaga to modify parts of the CSHELL fore-optics to accommodate our prototype instrumentation.  

Wavelengths are selected by adjusting the echelle grating central wavelength and a continuously variable filter for order selection that inadvertently introduces fringing at the $\sim$1-3\% flux level in our spectra.  Although this was a known limitation of CSHELL, we were unaware of the fringing at the time of commissioning our prototype.  The fringing from the filter adversely affected and delayed our data analysis pipeline development and resulting radial velocity precision.  Regardless, we were eventually able to incorporate a solution to the fringing into our analysis to obtain useful preliminary radial velocity measurements  (${\S}$6).

\subsection{Pre-Existing Calibration Unit}

The calibration unit of CSHELL is situated on top of the cryogenic entrance window (Fig 2). The calibration unit consists of a rectangular metal enclosure with a pass-through cavity approximately 12x6x7 inches.  The $\sim$2-inch diameter f/37.5 converging beam from the secondary mirror passes through this cavity along the longest dimension.  A fold mirror enters this cavity approximately half-way along the beam travel to inject calibration lamp light via an integrating sphere (flats, Xenon, Argon \& Krypton) that is mounted with the lamps on the other side of this cavity.  We have added an absorption gas cell into the space between this fold mirror and the calibration unit wall closest to the primary mirror, a volume approximately six inches on a side.  The gas cell instrumentation is discussed in detail in Plavchan et al. (2013) and Anglada-Escude et al. (2012).  We have added the fiber scrambler prototype in the space between the fold mirror and the calibration unit wall furthest from the primary mirror, a small volume only $\sim$3'' long in the direction of the telescope beam travel, $\sim$6" in one cross-dimension, and  $\sim$7" in the other cross-dimension before the fiber scrambler prototype protrudes from the calibration unit by less than $\sim$10''.

\begin{figure}[tb]
  \begin{center}
    \includegraphics[width=0.30\textwidth]{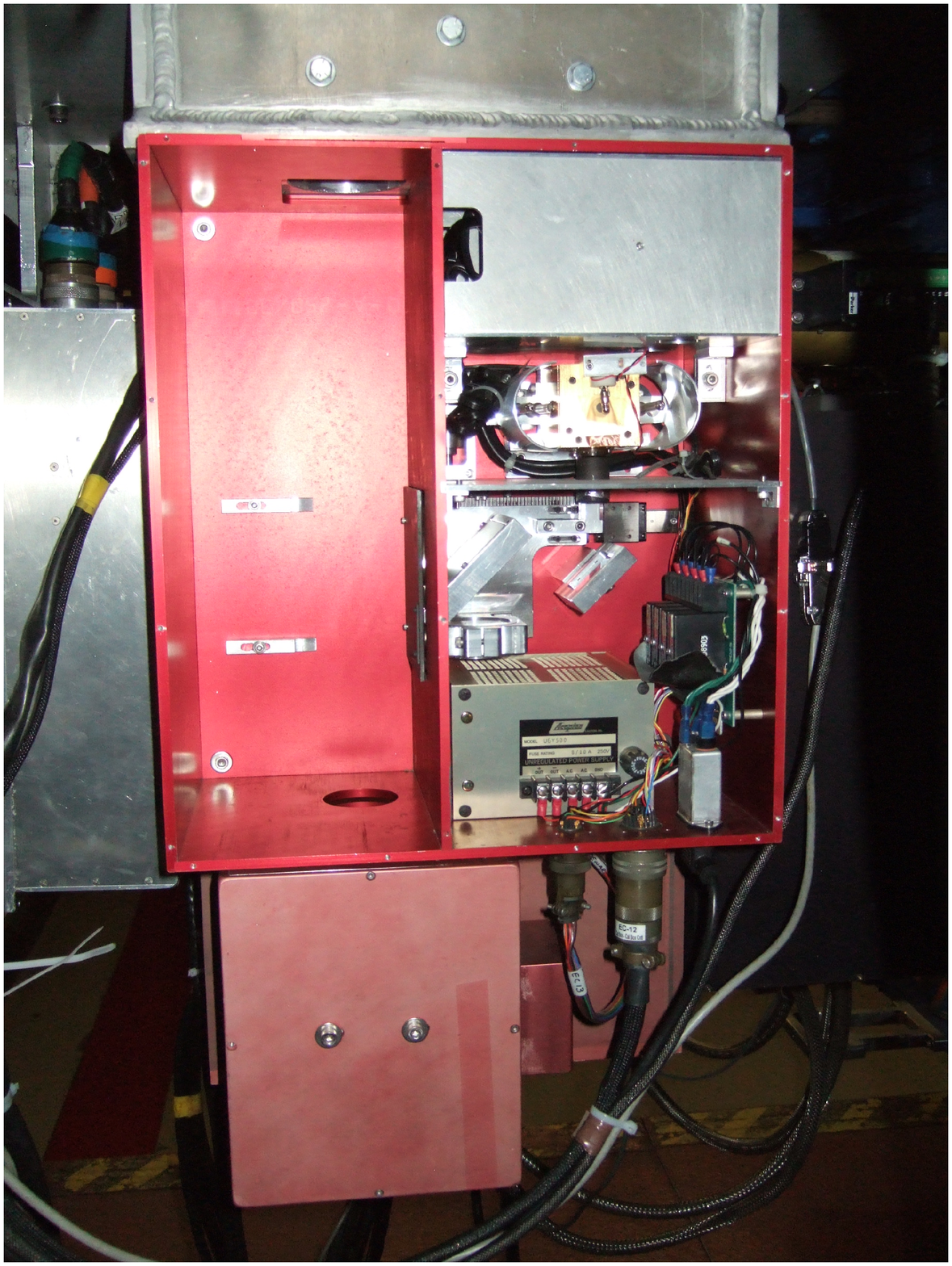}    
    \includegraphics[width=0.30\textwidth]{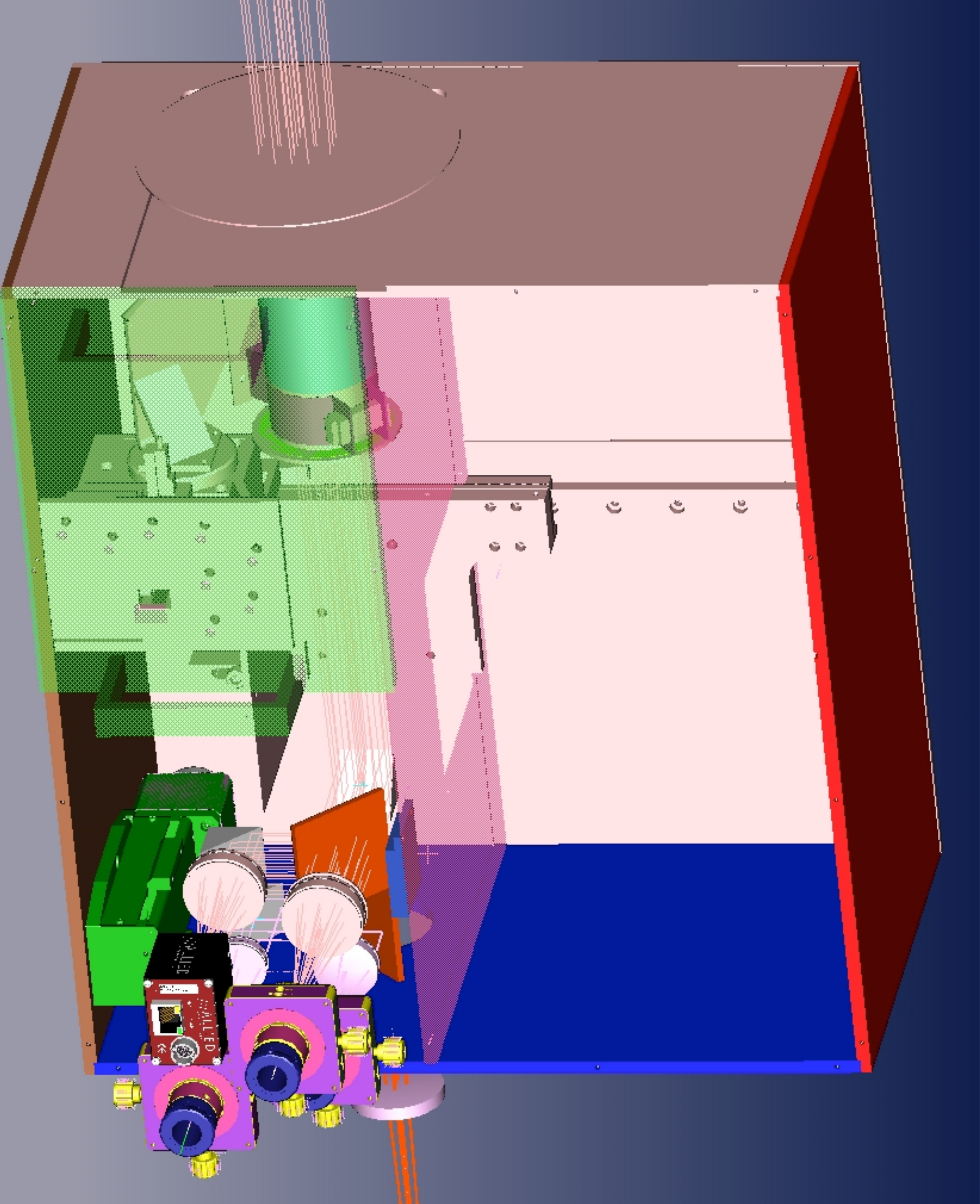}    
    \includegraphics[width=0.30\textwidth,clip=true,trim=20cm 0cm 0cm 0cm]{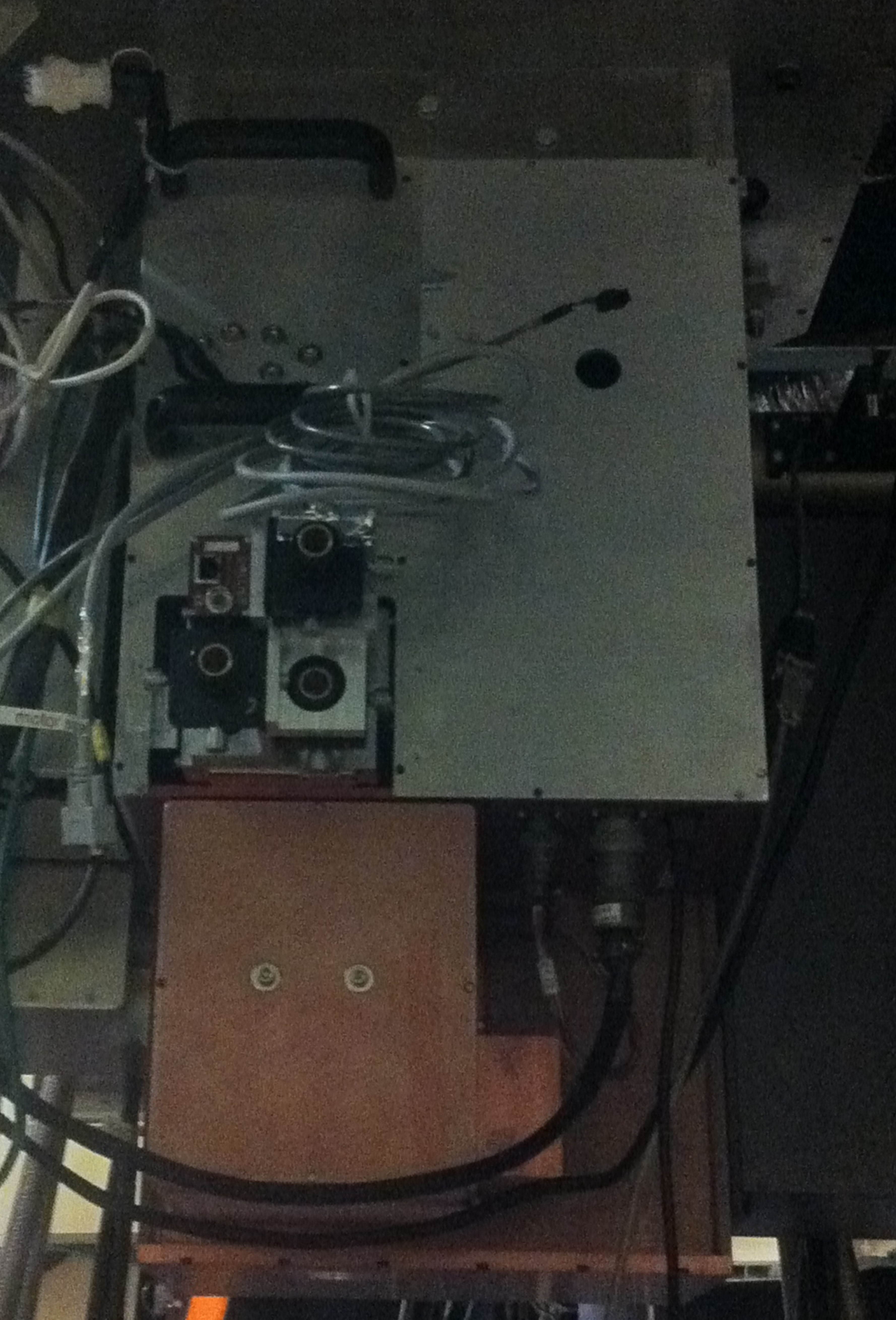}    
      \end{center}
      
  \caption{Left: CSHELL calibration unit with cover removed.  In this view, the primary mirror is towards the top of the image.  The beam converges from top to bottom on the left half of the calibration unit as seen in this image.  It is into this space we inserted the gas cell and fiber scrambler.   Middle: NX I-DEAS model view of the integrated gas cell and fiber scrambler with the CSHELL calibration unit.  The gas cell and mount is shown in the top left. The right hand side is empty as the calibration lamps, integration sphere and associated electronics are not shown.  The fiber scrambler optics and fiber mounts are in the lower left of this panel.  The control electronics box and control laptop are not shown, but are attached to a mounting plate to the left of the calibration unit.  The mount that secures the fiber scrambler optics to the linear stage (in green) is not shown for clarity.  Right: The completed gas cell and fiber scrambler integrated with the CSHELL calibration unit. Fibers and cables are not shown.  Three fiber chucks along with the alignment camera of the fiber scrambler are visible since the cover for the fiber scrambler is removed.  The handles are where the gas cell is inserted into the calibration unit above the fiber scrambler in this panel.  \label{fig:f2} }
\end{figure}

\subsection{Wavelength calibration source}

In this experiment, we use a methane gas absorption cell for a common optical path relative wavelength calibration.  A thorough description of the design, construction, and performance of this cell can be found in Plavchan et al. 2013 and Anglada-Escude et al. 2012.  Starlight passes through a 12.5 cm transparent cylinder filled with isotopic methane ($^{13}CH_4$) at 275 mb of pressure, which imprints a sharp set of absorption lines in the near-infrared, both in H and K band.  We use $^{13}CH_4$ rather than the more common telluric $^{12}CH_4$.  The absorption lines of the $^{13}CH_4$ gas cell are shifted in wavelength, so that the lines do not overlap with the variable telluric methane lines in the atmosphere. During normal operations, the cell is heated to 283 K, similar to a high ambient dome temperature, and maintained at +/-0.1 degree.  This level of temperature stability corresponds to errors below 1 m/s \cite{bean10}.

\section{Fiber Scrambler Optical Design}

\subsection{Optical Paths}

\begin{figure}[tb]
  \begin{center}
    \includegraphics[width=0.60\textwidth]{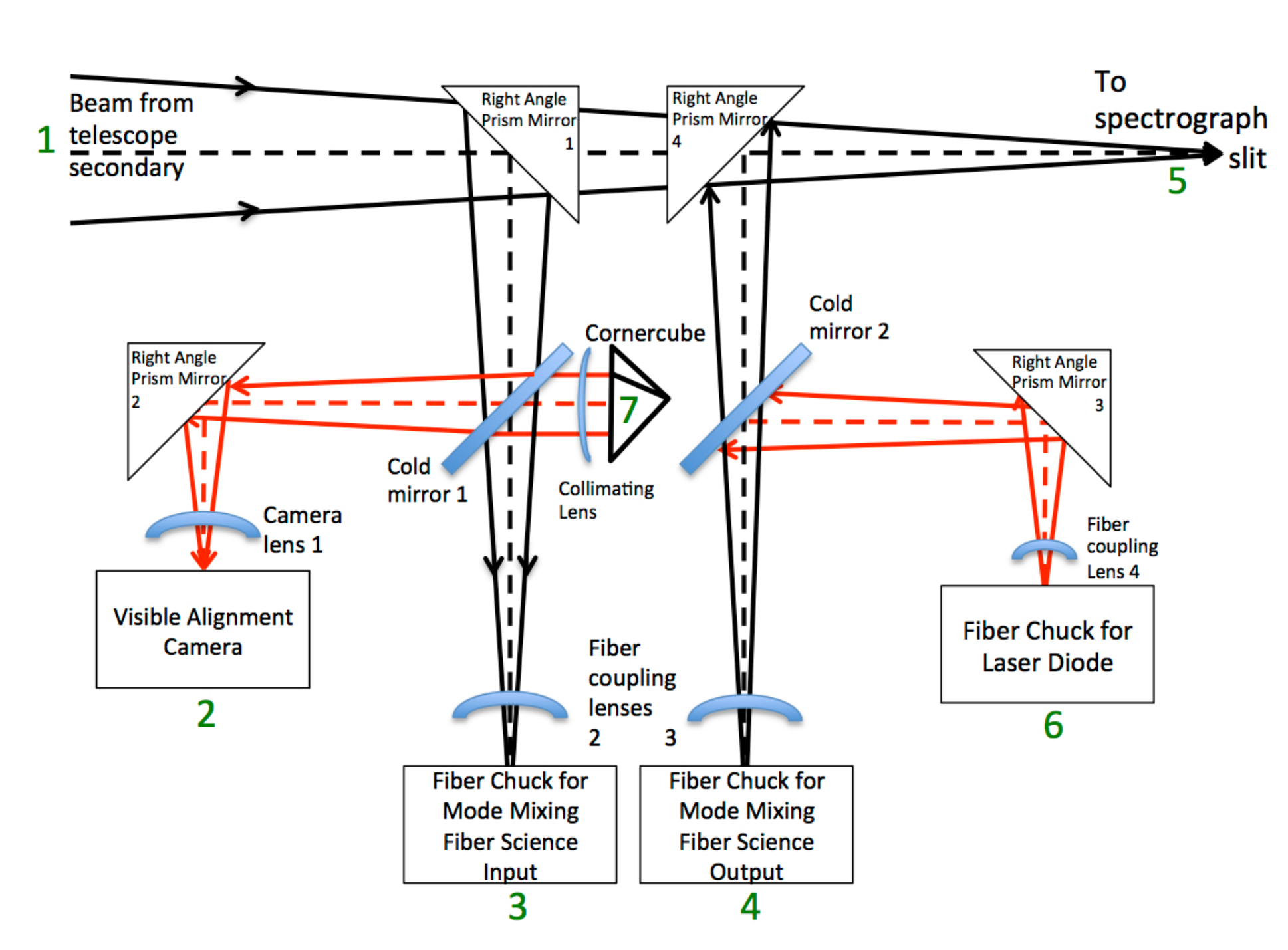}    
      \end{center}
  \caption{Optical path of fiber scrambler.  Starlight enters from the top left of the figure.  Instrument ``arms'' are numbered in green, and transmissive optics are in blue.  The left half of the figure is considered the ``input'' side and the right half is considered the ``output'' side.  On the input side, the telescope beam (arm 1) is re-directed by a pickoff fold mirror to the left cold mirror.  The left cold mirror acts as a dichroic and relays the input visible starlight (shown as red rays) to a second fold mirror and alignment camera (arm 2), and passes the input near-infrared light (shown as black rays) through to the fiber coupling lens and the fiber mounted at the left fiber chuck (arm 3).  On the output side, the near-infrared starlight exits the right fiber chuck and goes through a fiber coupling lens (arm 4).  This lens converts the diverging beam from the fiber back to the converging f/\# and beam diameter of the original telescope beam.  A second cold mirror dichroic on the right output side allows the infrared light to pass through towards the output fold mirror that redirects the near-infrared fiber output to the spectrograph slit and detector (arm 5).    Simultaneously, the cold mirror dichroic on the right directs laser light (arm 6) through the fiber ``backwards'' from the right fiber chuck to the left fiber chuck (arm 4 to 3).   The laser light exits the left fiber chuck, still traveling ``backwards'' compared to the near-infrared starlight.  The left cold mirror reflects the laser light to a corner cube (arm 7).  After the corner cube retro-reflection, a small percentage of the laser light is transmitted through the cold mirror, finally arriving at a focus at the visible alignment camera (arm 2).  \label{fig:f3} }
\end{figure}

In principle, a fiber scrambler only requires matching the telescope beam to the fiber input, and then relaying the fiber output onto the spectrograph slit input.  We developed a novel and more sophisticated design that has several distinct advantages (Figure 3).  There are three separate optical paths through the fiber scrambler totaling 7 optical ``arms'' -- one optical path is for the near-infrared science starlight, one for the visible starlight, and one for HeNe laser light used as a positional and focus reference.  Before entering the fiber scrambler, the near-infrared and visible starlight enter the calibration unit and pass through the gas cell at the native focal ratio of the telescope secondary mirror, converging at f/37.5 on the way to a Cassegrain focal plane (arm1; from top-left to top-right of Figure \ref{fig:f3}).  The first 1x1.4-inch pickoff 45$^\circ$ mirror is inserted into the telescope beam  $\sim$1 meter before the Cassegrain focus, and the near-infrared and visible starlight is redirected incident onto the first 2x2-inch diameter 45$^\circ$ dichroic.  The dichroic sends the visible starlight to the second 1x1.4-inch 45$^\circ$ fold mirror, the first 1-inch diameter achromat camera lens, and finally to the visible alignment camera in the focal plane of the lens (arm 2).

The near-infrared starlight passes through the first dichroic and is coupled by the second 1-inch diameter achromat lens that brings the beam to a focus at entrance to science fiber (NA=0.2; arm 3).  The starlight exits on the other end of the science fiber (N/A=0.2; arm 4) and passes through the third 1-inch diameter achromat lens that converts the diverging beam to a f/37.5 converging beam that matches the beam from the telescope secondary mirror.  With this setup, the spectrograph is fed with the same beam diameter and f/\# regardless of whether the fiber scrambler is inserted or not.  The near-infrared starlight exiting from the science fiber passes through a second 2x2'' 45$^\circ$ dichroic unaltered (save for a small throughput loss).  The near-infrared starlight finally re-enters the telescope beam via the fourth 1x1.4-inch 45$^\circ$ fold mirror (arm 5).  

Separately, a laser diode sends HeNe laser light out of a single-mode fiber to a 1/2-inch lens that converts the laser light to a diverging f/37.5 beam (arm 6). The laser light continues to the third 1x1.4-inch 45$^\circ$ fold mirror that redirects the laser light to the second dichroic.  The dichroic reflects the laser light ``backwards'' in the opposite direction of the near-infrared star-light coming out of the science fiber.  The laser light is focused by the third lens and enters the science fiber on the ``output'' end of the fiber (arm 4) and exits on the ``input'' end (arm 3).  The laser light, after passing through the second lens is then reflected by the first dichroic onto a collimating lens and finally a cornercube retroreflector (arm 7).  The retroreflector reverses the laser light back to the collimating lens and dichroic.  A small percentage of the laser light is transmitted through the dichroic (most is reflected back to the fiber), after which it is focused onto the visible camera along with the visible starlight (arm 2).

To summarize, the paths of the near-infrared starlight are arms $1\rightarrow3\rightarrow4\rightarrow5$, visible starlight $1\rightarrow2$, and laser light $6\rightarrow4\rightarrow3\rightarrow7\rightarrow2$.  The net result is that arms 1 and 5 match the original telescope beam when the fiber scrambler is not in place.  Arm 7 containing the retro-reflector and collimating lens is short, and sticks out perpendicular to the other optical paths.  Because of the second and third 45$^\circ$ fold mirrors, arms 2,3,4 and 6 are parallel to one another and can be arranged in a compact 2x2 grid.  Two of these arms (2 and 6) are visible wavelength arms for the camera and laser injection, and the other two arms propagate near-infrared science light and laser light in opposite directions through the science fiber (arms 3 and 4).   

\subsection{Cornercube Retro-reflector}

The use of the retroreflector and laser are novel in our fiber-fed design.  It enables the simultaneous image acquisition of the star and laser-illuminated fiber at a wavelength other than the desired science wavelength.  Because we use a retroreflector and not a flat mirror, there is no misalignment of the flat mirror and reflected laser beam.  The relative positions of the laser light and the visible star-light on the camera can be used as a convenient proxy for placing and guiding the near-infrared star-light onto the input science fiber tip.  All that is needed to establish and maintain the near-infrared starlight on the fiber tip is to align the two images on the camera by small pointing adjustments to the telescope.   We neglect small chromatic differences due to the wavelength dependent index of refraction of air.

In practice, we found that initially positioning the star on the fiber took a few seconds, and the external IRTF telescope guider was accurate enough for most normal observations to maintain the star on the tip of the fiber to within $\sim$0.''1.  In the cases of the external IRTF guider glitches or failure, we were able to manually guide the telescope with ease with the visible alignment camera, as the visible camera was operated at a high frame-rate compared to the long exposures of the spectrograph.  There are a number of advantages to this guiding system.  It eliminates the significant overhead time and tedium of blindly moving the star onto the fiber tip after a slew and/or a guider glitch, and greatly simplifies the alignment adjustment to maximize the starlight throughput coupling to the fiber.  It adds a second method of guiding in the case of a lack of guide stars or an external guider failure.  Furthermore, it does this without using any of the starlight at the science wavelengths; it uses light that would be discarded by the spectrograph.  As a caveat, the first cold mirror dichroic does have multiple reflections of the laser light due to its non-zero thickness.  Thus, care must be taken to determine which reflection is the ``real'' one (the brightest), and a thicker dichroic helps displace these multiple reflections.  

\subsection{Individual Optics}

The two dichroics are low-cost ($<$\$100) cold mirrors that show excellent transmission and reflectance for our desired wave-bands, with a transition from reflection to transmission at 0.7 $\mu$m as shown in Figure 4.  Two additional and more expensive off-the-shelf dichroics have also been tested with slightly better near-infrared transmission and 3-5 mm thickness.  The four lenses are all off-the-shelf achromats that are near-infrared anti-reflective coated with a standard coating for excellent transmission at 1.6 $\mu$m and slightly less transmission at 2.3 $\mu$m and in the visible.  The four 45$^\circ$ fold and pickoff mirrors are all off-the-shelf silver coated right angle prism mirrors.    The single retroreflector is off-the-shelf and 1/2'' in diameter.  A bike reflector consisting of many small plastic retroreflectors was tested, but produced a spatially incoherent reflected laser beam of many misaligned retro-reflections.  The fibers are described in ${\S}$3.4.  The visible acquisition camera is an Allied Vision GC1290 Megapixel CCD with ExView Sensor that requires no active or passive cooling.  No filters are applied to the camera optics and the read noise is negligible, with the camera readily able to detect our target stars as faint as V=12 with integration times suitable for guiding ($<$1 s).  All optics were set up and tested on a lab bench prior to integration with the machined mount described below.

\begin{figure}[tb]
  \begin{center}
    \includegraphics[width=0.48\textwidth,clip=true,trim= 2cm 2cm 2cm 2cm]{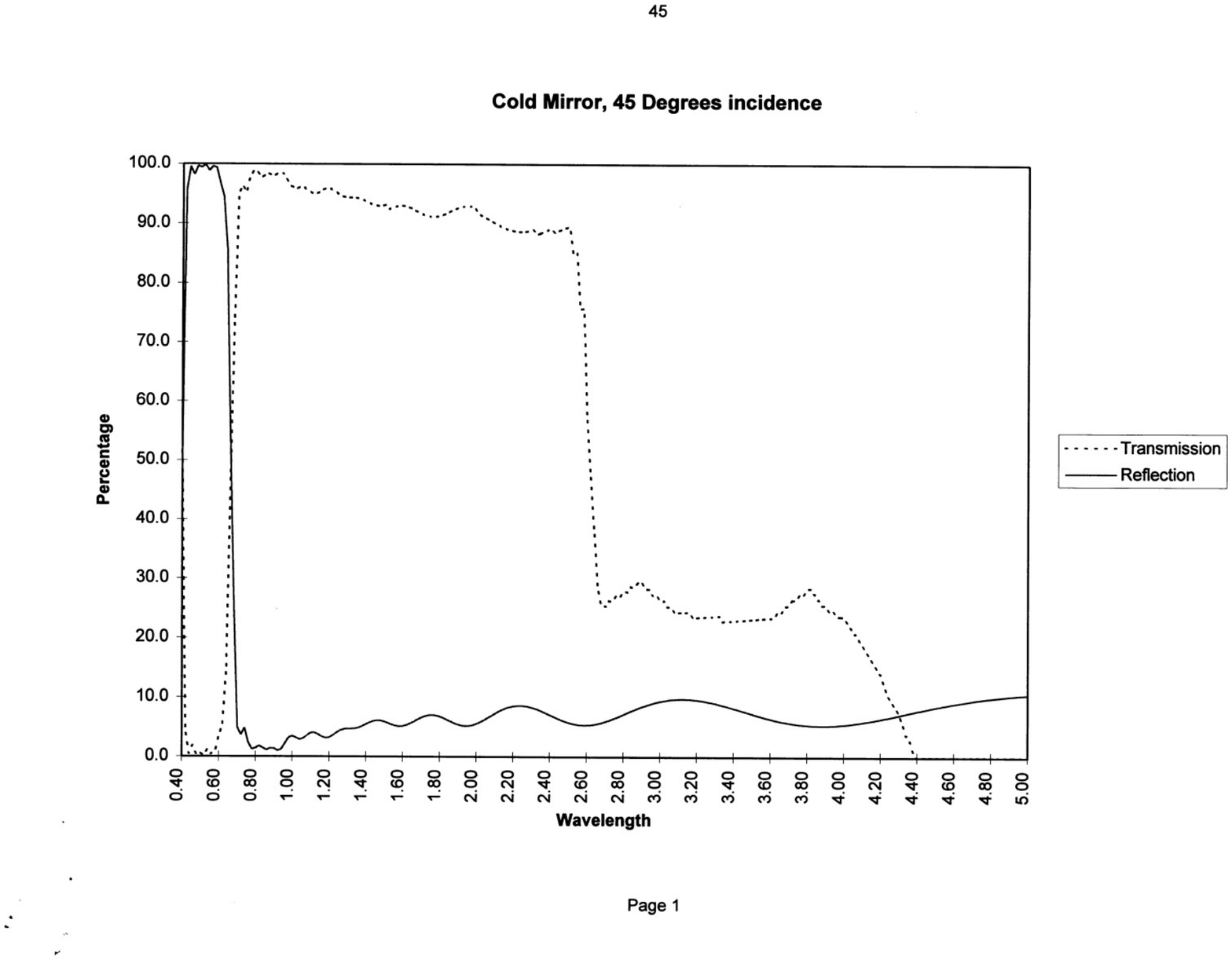}    
    \includegraphics[width=0.48\textwidth,clip=true,trim= 0cm 2cm 2cm 2cm]{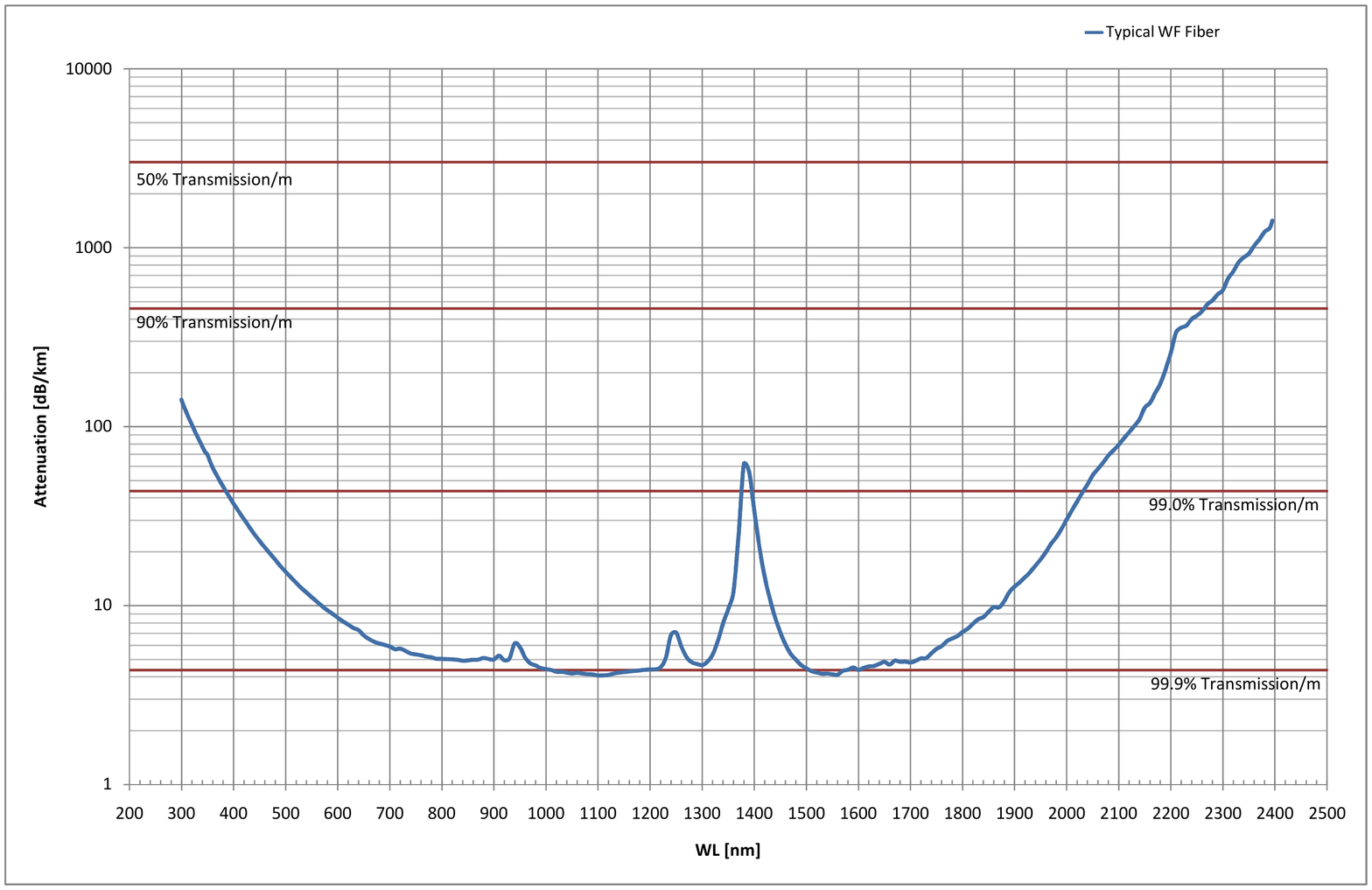}    
      \end{center}
  \caption{Left: 45$^\circ$ dichroic reflection and transmission curves used to split the visible and near-infrared pathways in the fiber scrambler.  Wavelength is on the horizontal axis, and the percentage of flux transmitted/reflected on the vertical axis.  Right: Fiber attenuation spectrum for a typical fiber such as those used in this paper.  The horizontal axis is the wavelength, and the vertical axis is the attenuation not including reflection at the entrance tip of the fiber.  As can be seen, the attenuation is minimal in the H-band, but rises steeping in the K-band to render observations at K-band effectively useless in terms of throughput. \label{fig:f4} }
\end{figure}

One major drawback of using the fiber scrambler with CSHELL is that CSHELL has a chromatic lens after the calibration unit.  This lens is used to speed up the converging f/37.5 beam to $\sim$f/15 to reach a faster Cassegrain focal plane at the slit.  Thus, any changes to the wavelength of observations requires a manual refocussing of the fiber output onto the spectrograph slit to account for the chromatic aberration of this lens.

\subsection{Fibers}

We solicited quotes from several fiber vendors that produce non-circular fibers, including CeramOptec, Le Verre Fluore, Polymicro and Molex.  The cost of the fiber (for any length of $\sim$1-20 m) was primarily determined by whether or not the vendor had already made a fiber of the desired core geometry and diameter for another customer and thus had a $\sim$km long spool of the fiber available for purchase off-the-shelf.  If the desired fiber has not already been manufactured, the cost to manufacture the fiber is typically \$20-50k in USD.  Off-the-shelf fibers cost $\sim$\$1-5k per piece.  As mentioned in ${\S}$1, many non-circular core fibers are made for other applications in applied pump laser physics.

From CeramOptec, we purchased a number of Optran NCC fused silica core fiber assemblies -- octagonal and square core fibers of diameter 200 $\mu$m and lengths 1,8 and 10 m, 50x100 $\mu$m rectangular core fibers of length 1 and 10 m, and 50 micron octagonal core fibers of length 1 and 10 m.  All fibers were FC connector flat terminated, with a numerical aperture NA=0.2--0.22.  The transmission loss of the typical fused silica fiber is shown in Figure 4, with excellent transmission in the H band at 1.6 $\mu$m. Due to the low-cost nature of this prototype, we were not able to manufacture a fiber with a core material such as fluoride glass that has low transmission loss at $>$2 $\mu$m in wavelength, and consequently we purchased the 1 meter silica samples for testing observations at K band.  The imaged fiber tip core geometries are shown in Figure 5.

\begin{figure}[tb]
  \begin{center}
    \includegraphics[width=0.80\textwidth]{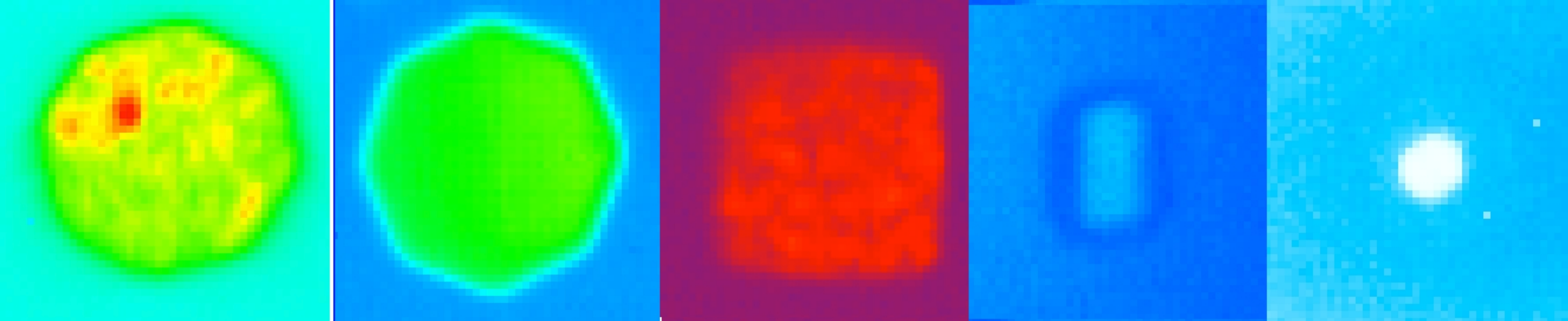}    
      \end{center}
  \caption{ Near-infrared images of non-circular core fibers taken of standard stars with CSHELL at IRTF at 1.65 $\mu$m.  From left to right: 1 m non-agitated 200 $\mu$m octagonal core, 10 m agitated 200 $\mu$m octagonal core, 10 m 200 $\mu$m square core (contrast altered to show modal noise; ${\S}$3.6), 50x100 $\mu$m rectangular core, and 50 $\mu$m octagonal core fibers respectively.  The two 200 $\mu$m octagonal core fibers are approximately the same stretch and show significant modal noise for the 1 m fiber compared to the agitated 10 m fiber. For the rectangular fiber, the cladding (material around the core) is propagating some starlight.  The sides of the 50 $\mu$m octagonal core fiber are only marginally resolved by the CSHELL imager. \label{fig:f5} }
\end{figure}

\subsection{Image Slicer}

The initial design of our fiber scrambler called for the use of the 200 $\mu$m diameter fibers, corresponding to $\sim$3'' on the sky.  Due to the mechanical design limitations placing the science fiber output $\sim$1 m away from the slit plane, this fiber was severely over-sized for the 0''.5 slit.  Thus, to improve the throughput we designed an image slicer based upon the Yale two clocked mirror image slicer \cite{chiron,chiron2,chiron3} shown in Figure 6.    Instead of the two clocked mirrors with an air gap, we designed a single circular flat substrate that we would have coated via photolithography with two clocked reflective coatings on either side.  However, the image slicer design proved too costly to implement, and too difficult to fit in the space available for the fiber scrambler.  Thus, instead of adding the image slicer, we went with the smaller 50x100 $\mu$m rectangular and 50 $\mu$m octagonal fibers to lower the slit throughput losses.  

With the image slicer design discarded, a slit became necessary to cut down the fiber diameter to avoid sacrificing a loss in spectral resolution.  Rather than using the spectrograph built-in slits, we also purchased and mounted two aluminum slits (of 10 and 25 $\mu$m diameters) to fix the position of the slit with respect to the fiber tip.  However, internal reflections between the slit and tip of the fiber (in close proximity) produced ghost images on the detector that rendered the spectra unusable for precision radial velocities.  Thus, we used the spectrograph slit for our observations.  With sufficient space, the formation of a second image plane of the fiber tip at which to place the slit would have eliminated the ghost images.

\begin{figure}[tb]
  \begin{center}
    \includegraphics[width=0.40\textwidth]{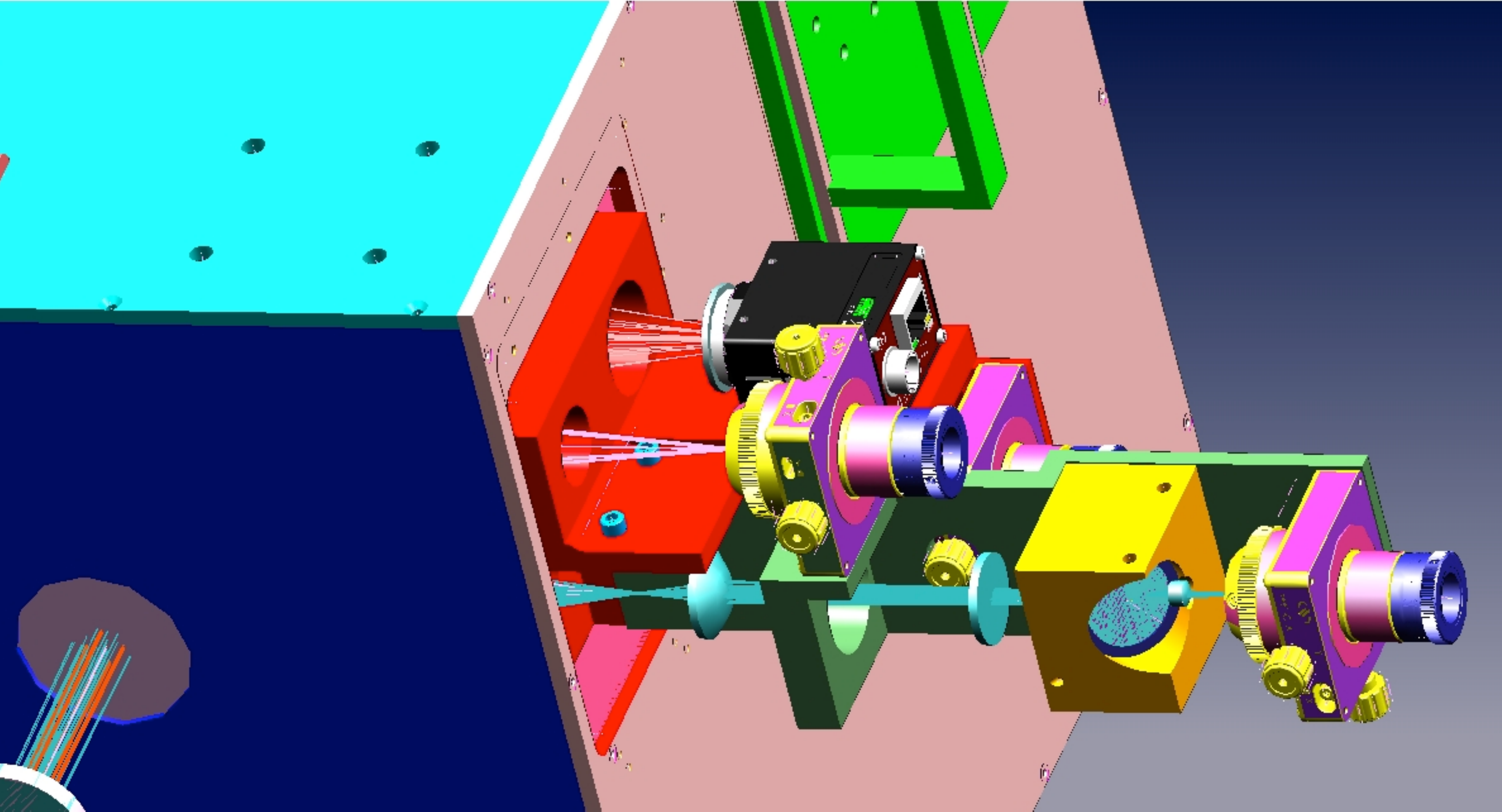}    
      \end{center}
  \caption{NX I-DEAS model of the Image slicer.  The slicer is located in the yellow cube in the lower right, moving the science fiber chuck several inches further out and adding two additional lenses (mounts not shown for clarity) to decrease the beam speed before slicing.  The main fiber scrambler mount is shown in red with the extension in green, and the visible alignment camera and laser diode fiber chucks are visible in black and purple with yellow knobs respectively.  At the top of the figure the green handle is for the gas cell.  \label{fig:f6} }
\end{figure}

\subsection{Fiber Modal Noise}

Fiber modal noise is the varying spatial intensity across the fiber tip at the exit of the fiber.  It is due to the interference of the finite number of different modes of propagation in a fiber.  The output depends on the input illumination, the core fiber geometry, and the path length through the fiber.  It is variable with changes in the optical path, for example due to temperature and stress in the fiber.  In Figure 7 we show sample visible lab tests with a HeNe laser of some of our fibers.  

\begin{figure}[tb]
  \begin{center}
    \includegraphics[width=0.40\textwidth]{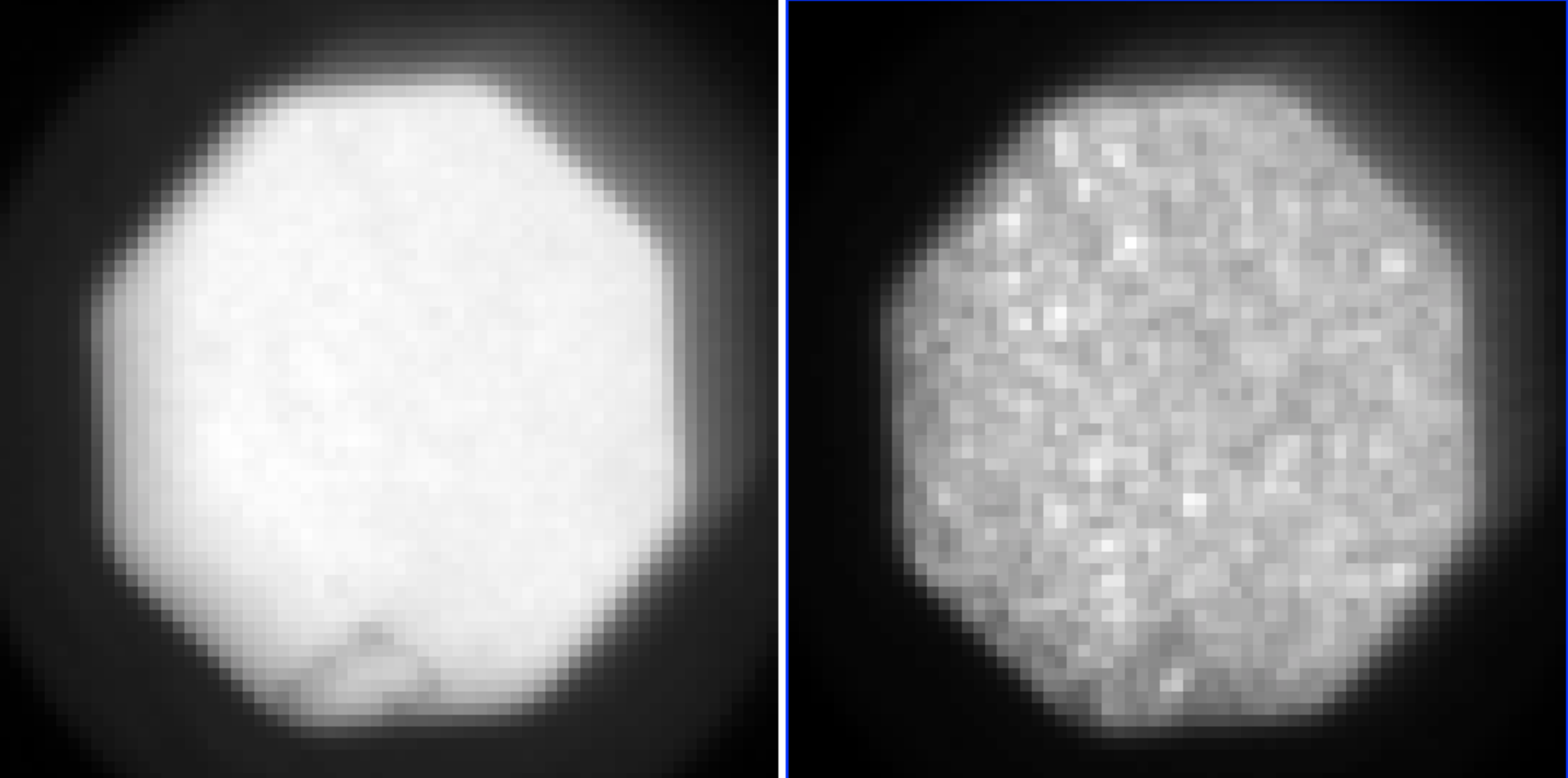}    
    \includegraphics[width=0.18\textwidth]{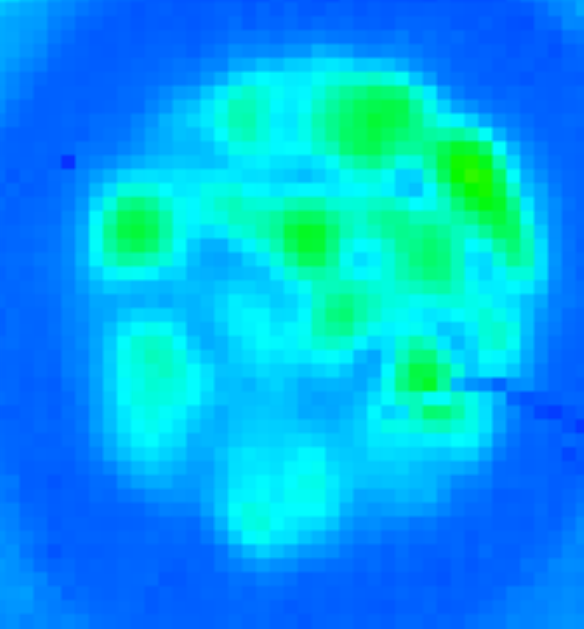}    
      \end{center}
  \caption{ From left to right: 10 m agitated fiber output illuminated with a HeNe laser; 10 m non-agitated fiber output illuminated with a HeNe laser; 2.3 $\mu$m image of a 1 m non-agitated fiber illuminated by a standard star.  All three panels are near-field images of the 200 $\mu$m octagonal core fiber tips.  The first two figures were taken in the lab optical bench, while the last was taken with CSHELL at IRTF. Modal noise dominates the CSHELL image, and clearly exhibits the challenge of using fibers for near-infrared spectroscopy.  Modal noise is still readily apparent at visible wavelengths in the 2nd panel without agitation.  However, there are many more modes, and the modes are more pronounced in the visible due to the coherent monochromatic nature of the laser light.  Agitation in the first panel readily equalizes the individual mode output illumination (${\S}$4.2).\label{fig:f7} }
  \end{figure}

The stability of the output illumination from the fiber directly translates into the stability of the spectrograph line spread function and consequently the obtainable radial velocity precision.  The ratio of mean intensity to the standard deviation can be thought of as a signal-to-noise ratio, dependent on the number of speckles and any vignetting of the fiber tip.  This undesirable variability is referred to as ``modal noise'' by the photonics industry, and this is the traditional definition of modal noise \cite{epworth79}.

The approximate number of spatial modes excited in a multimode fiber uniformly illuminated by a coherent, monochromatic light is given by 

\begin{equation}
M = \frac{1}{2}\left( \frac{\pi d \ \mathrm{NA}}{\lambda}\right)^2
\end{equation}

\noindent where $M$ is the number of modes, $d$ and NA the core diameter and numerical aperture of the fiber, and $\lambda$ the wavelength of light \cite{goodmanrawson1981,lemke11}.  

When illuminating a fiber entrance by a coherent monochromatic wavelength source, we can derive the one-dimensional slit illumination function for the spectrograph by integrating the near-field image of the fiber exit across one direction (e.g. collapsing the panels in Figure 7 along the vertical direction). We can then vary the illumination of the fiber input, and measure the variation in the output illumination, by moving the illumination source across the fiber entrance (or by allowing for the PSF variations of starlight).   Finding the configuration that minimizes amount of variability of this simulated slit illumination function corresponds to the optimal setup for precision radial velocity work.  Since we did not have a near-infrared detector to work with in the lab, we performed some tests on sky with CSHELL, and some with the HeNe laser at visible wavelengths in the lab.  The resulting slit illumination functions and their variability from the HeNe lab tests are shown in Figure 8.  The analysis of the on-sky data is ongoing and will be reported in Bottom et al. (in prep.).  This approach assumes the absence of a slit and perfect spectrograph optics -- variable far-field illumination of spectrograph optics with inhomogeneities such as imperfections in the echelle ruling impact the LSF. The importance of the far-field illumination stability and analysis thereof is presented in Spronck et al. (2012a,b).

\begin{figure}[tb]
  \begin{center}
      \includegraphics[width=0.2\textwidth,clip=true,trim=4cm 0cm 4cm 0cm]{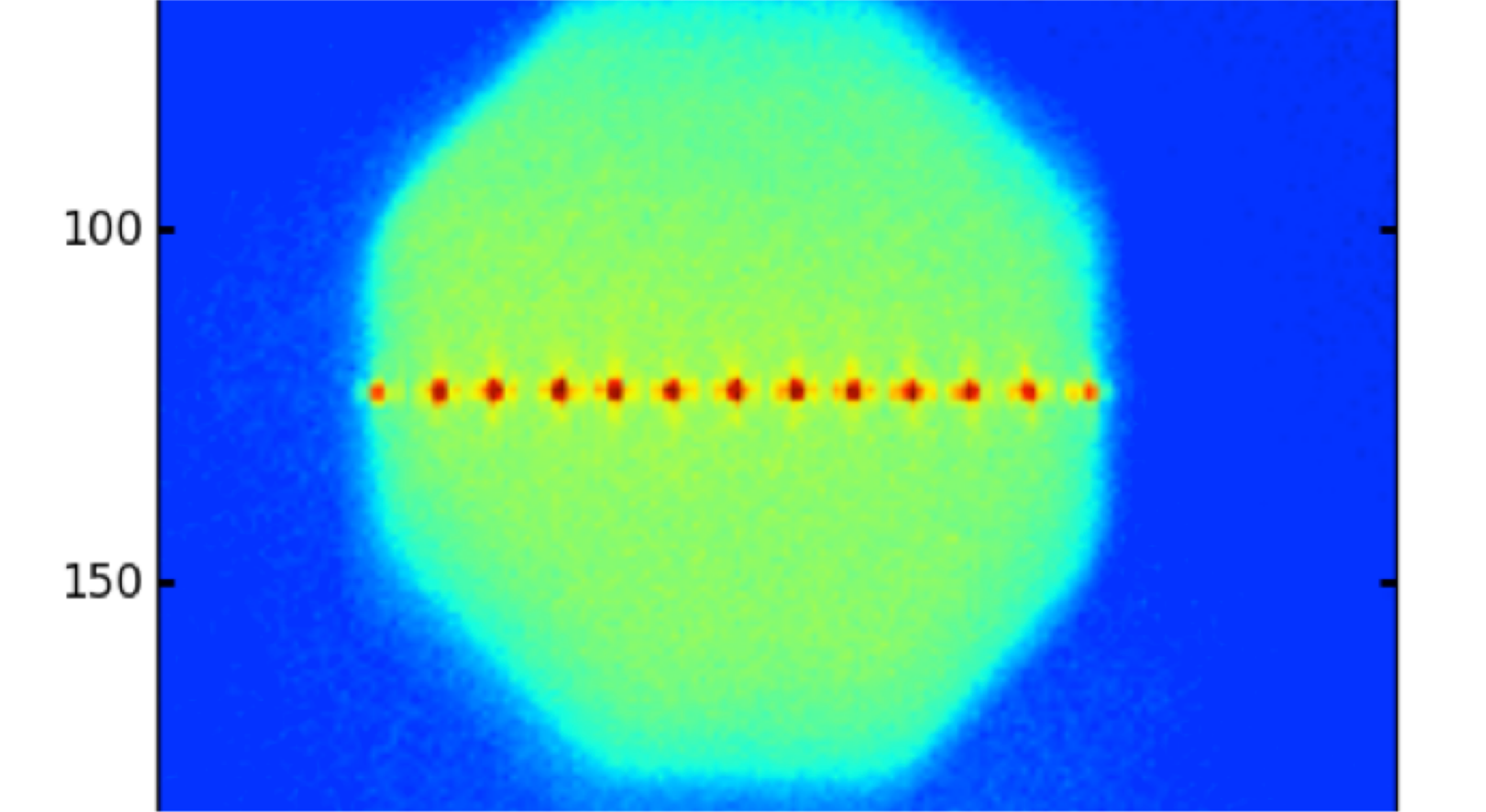}    
    \includegraphics[width=0.29\textwidth]{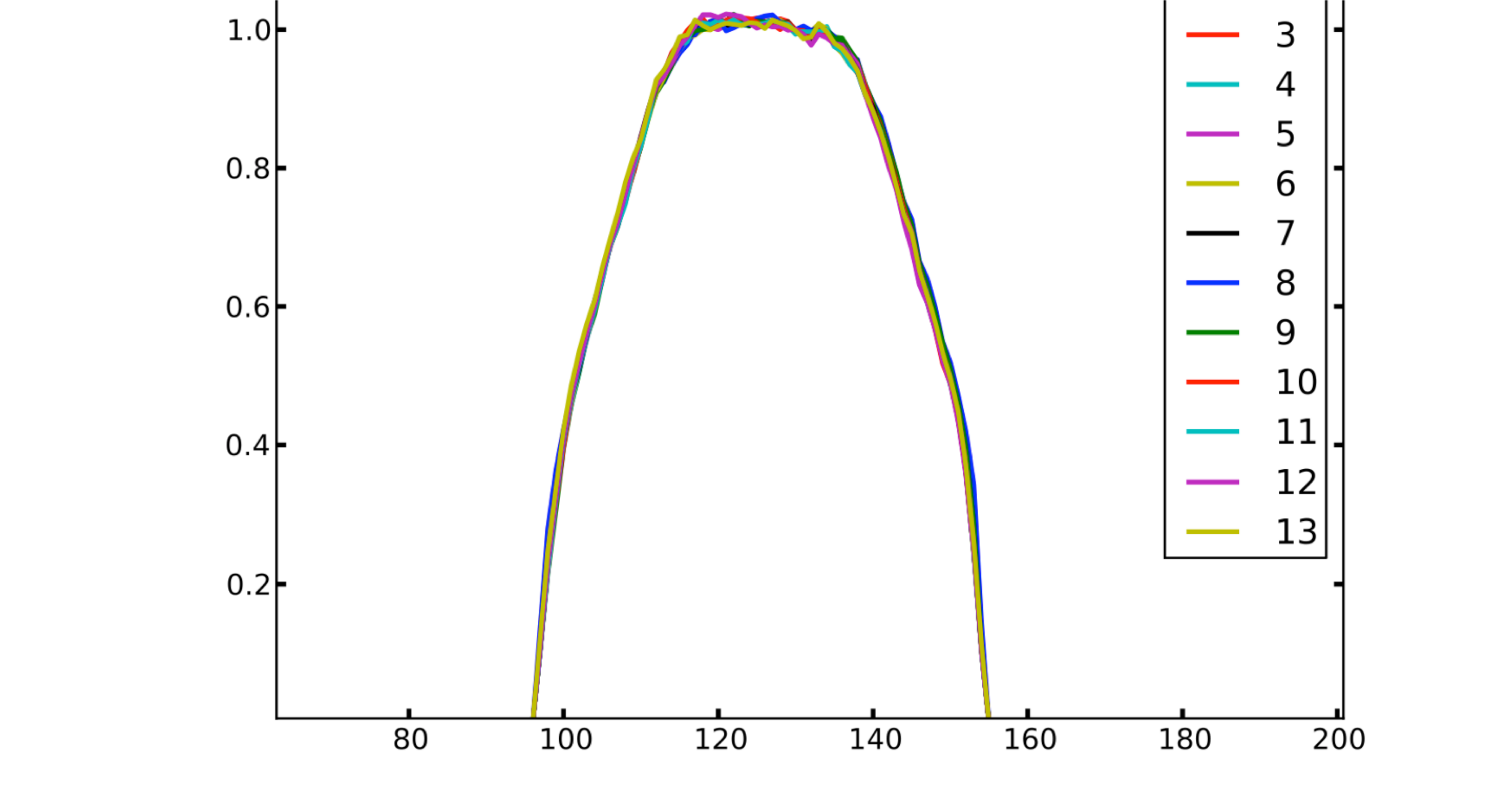}    
    \includegraphics[width=0.49\textwidth]{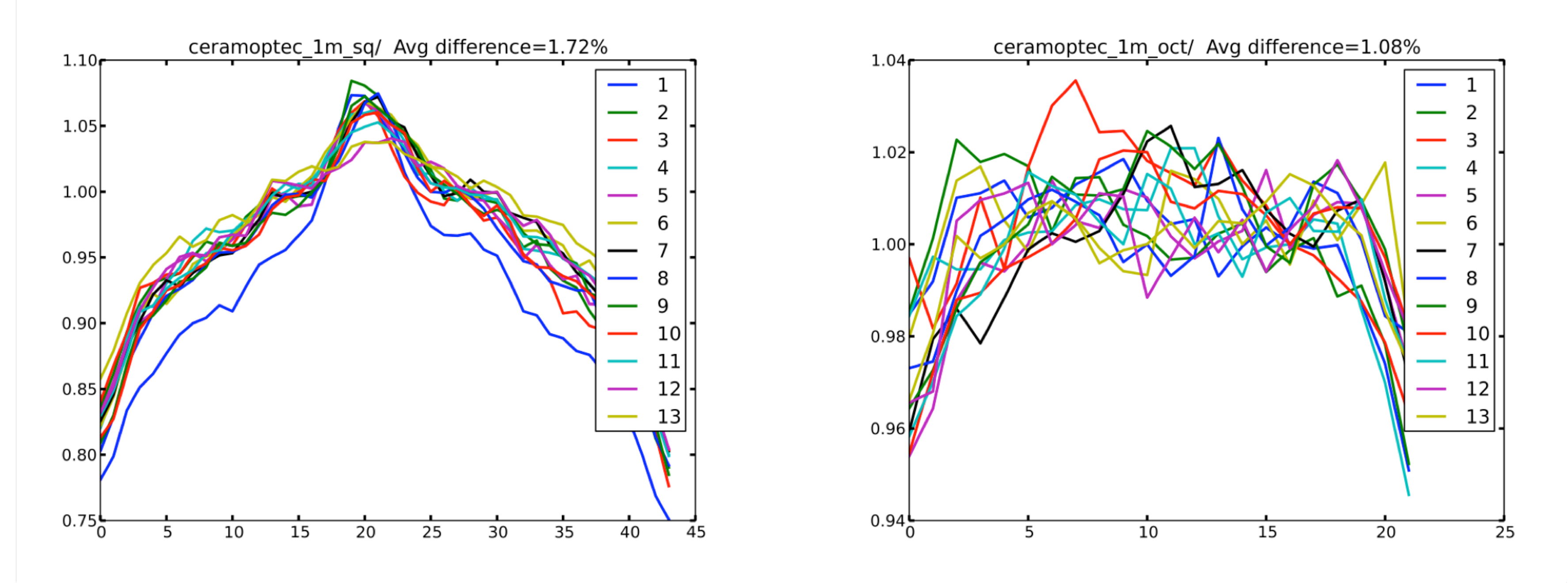}    
    \includegraphics[width=0.49\textwidth]{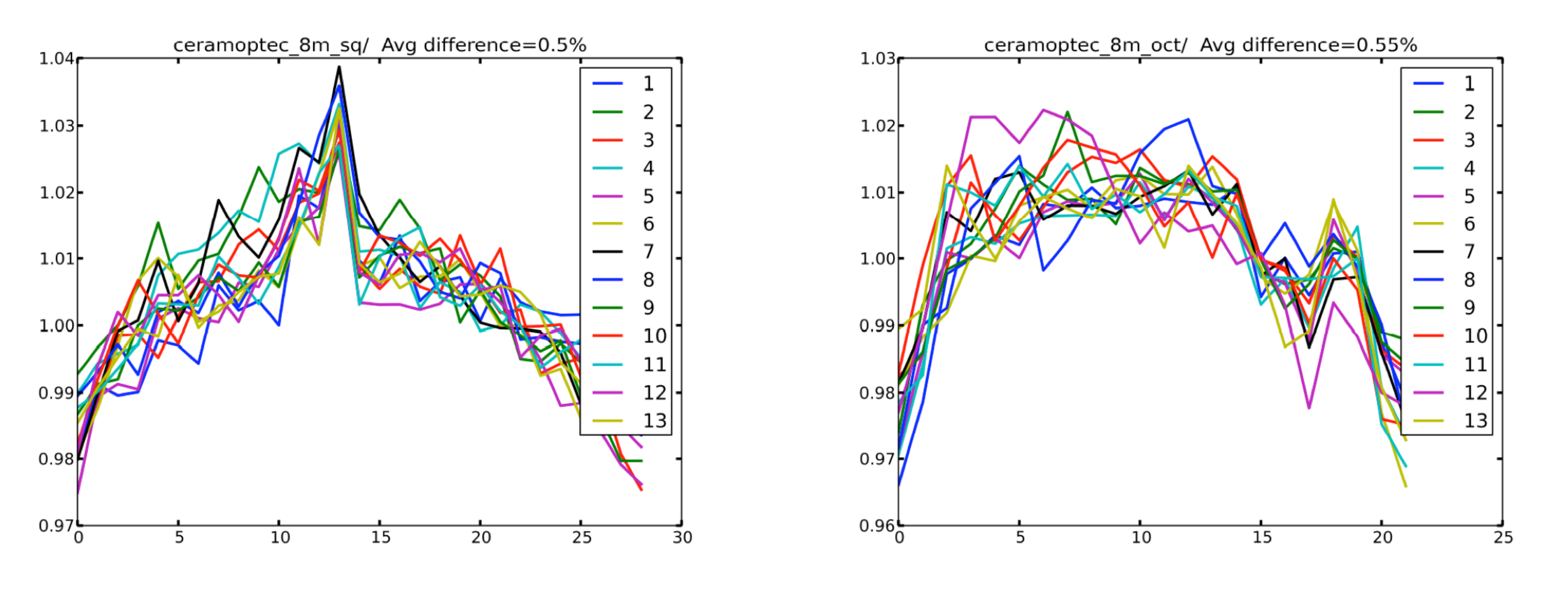}    
    \includegraphics[width=0.49\textwidth]{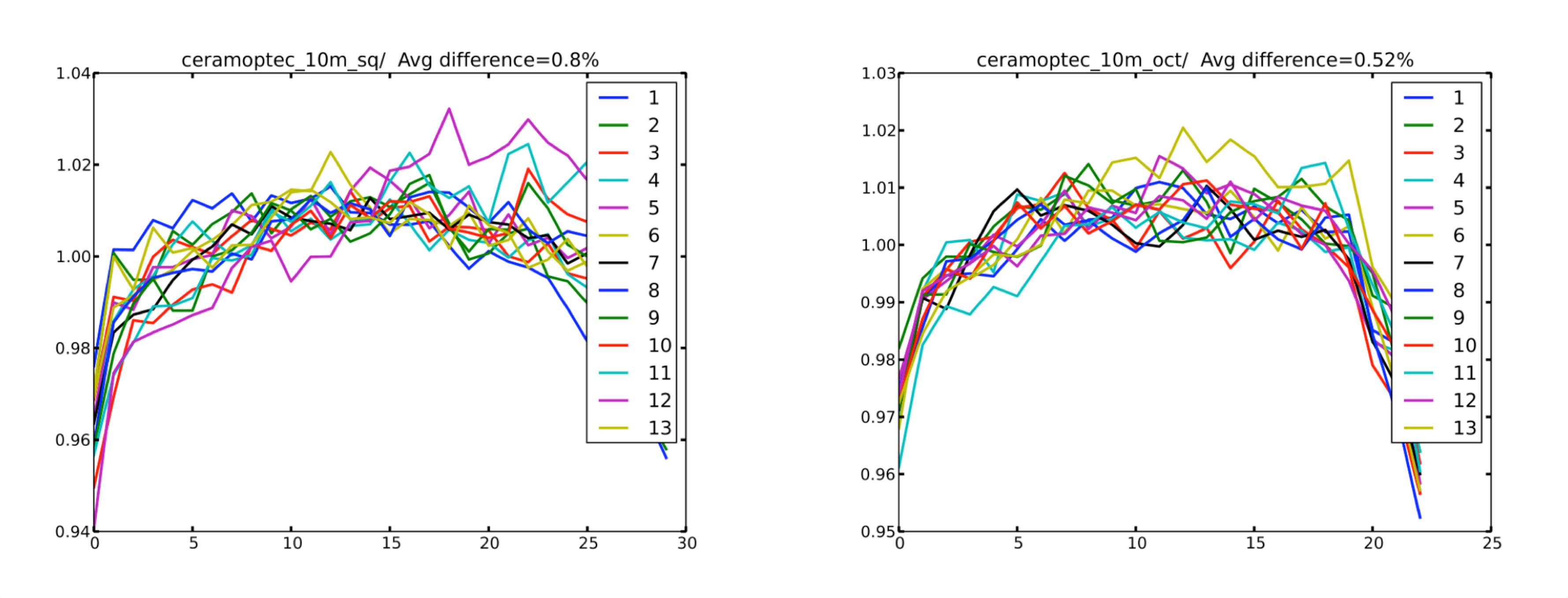}    
      \end{center}

  \caption{First row, left: Red HeNe laser spots reflecting off of the 10 m fiber input, overlaid from thirteen separate images, showing the thirteen different illumination positions of the laser w/r/t to the fiber tip ($\sim$17 $\mu$m spacing); First row, second from left: HeNe wavelength simulated slit illumination functions for the 8 meter octagonal core fiber for 13 different input illuminations.  The 13 different input illuminations are shown in the third panel of Figure 7.   The horizontal axis is pixels, and vertical axis is normalized intensity.  First row, middle: 1 meter long 200 $\mu$m diameter square core fiber simulated slit illumination functions for the same set of input illuminations, and zoomed on the vertical and horizontal axes to the peak of the curves to more clearly show the variations.  1st row, right: 1 meter long 200$\mu$m diameter octagonal core fiber simulated slit illumination functions.  Second row, left two panels: The equivalent slit illumination functions for 8 meter long fibers.  Second row, right two panels: The equivalent slit illumination functions for 10 meter long fibers.  The average deviation is indicated for each fiber core geometry and fiber length, with the conclusion that the octagonal fibers and longer fibers offer better scrambling.  The square fibers tend to exhibit linear spatial modes along the central axis (since it is a rectangular waveguide, as exhibited by the central ``peak'' in the middle of the fiber output illumination), which is indicative of poorer mode scrambling compared to the octagonal fibers. For both the square and octagonal fibers, greater mode scrambling is achieved compared to circular core fibers in response to input illumination variations, as has been demonstrated elsewhere\cite{spronck2}. No agitation was applied to the fibers.  \label{fig:f8} }
\end{figure}

One of the results of our analysis is that the full benefits of the non-circular core fiber geometries are not realized for fibers as short as 1 m. The longer non-circular core fibers offer superior spatial mode scrambling compared to circular fibers (not shown), which can have relatively dramatic changes in the output illumination in response to input illumination variations (e.g. $>$10\%).  By comparison, $<$2\% LSF variation is seen for  the 1 meter non-circular core fibers, and $<$1\% for the longer non-circular core fibers in the visible.  The second result is that agitation in addition to the non-circular core geometry further enhances the stability of the output illumination as can be seen in the first panel of Figure 7 (${\S}$4.2).  This has already been demonstrated with non-circular core fibers at visible wavelengths \cite{spronck2}.

\section{Fiber Scrambler Mechanical, Electronic and Software Design}

\subsection{Mechanical Design}

Herein we present the mechanical design of the fiber scrambler, including the mechanical agitation of the fiber.  The available physical space for the fiber scrambler ($\sim$ 3 $\times$6 $\times$18 in$^3$) places severe design constraints on the fiber scrambler optics and the motor mechanism to move the fiber scrambler in and out of the telescope beam.  The length of the fiber scrambler along the direction of the telescope beam travel -- $\sim$3 in -- is limited on one end by the exit window and wall of the calibration unit, and on the other end by the descending fold mirror for the calibration lamps.  Since the telescope beam travels in close proximity along two walls of the calibration unit (top of Figure 3 and the rear wall and right wall in the left panel of Figure 2), the fiber scrambler optics are all constrained to be on one side of the pickoff mirrors, and the options are limited for off-the-shelf stages.  

We replaced the cover for the calibration unit with a cover with a hole to permit the fiber scrambler to protrude out of the calibration box (foreground of Figure 2).   For practical (and cost) reasons regarding the protrusion of the fiber scrambler from the calibration unit, we eliminated the use of an image slicer from our design as discussed in ${\S}$3.4.  Additionally, the use of two fold mirrors are employed to arrange arms \#2 and 6 underneath arms \#3 and 4 as shown in Figures 2 \& 3 such that the four arms are arranged in a compact 2x2 grid.  There is also some vignetting of the off-axis beam to meet the space requirements, and to make use of off-the-shelf one inch optics.  This was an acceptable design choice given that we are only interested in the on-axis throughput of the starlight to the fiber and from the fiber.  The final mechanical design in shown in Figures 9 \& 10.

\begin{figure}[tb]
  \begin{center}
    \includegraphics[width=1\textwidth,clip=true,trim=0cm 6cm 0cm 6cm]{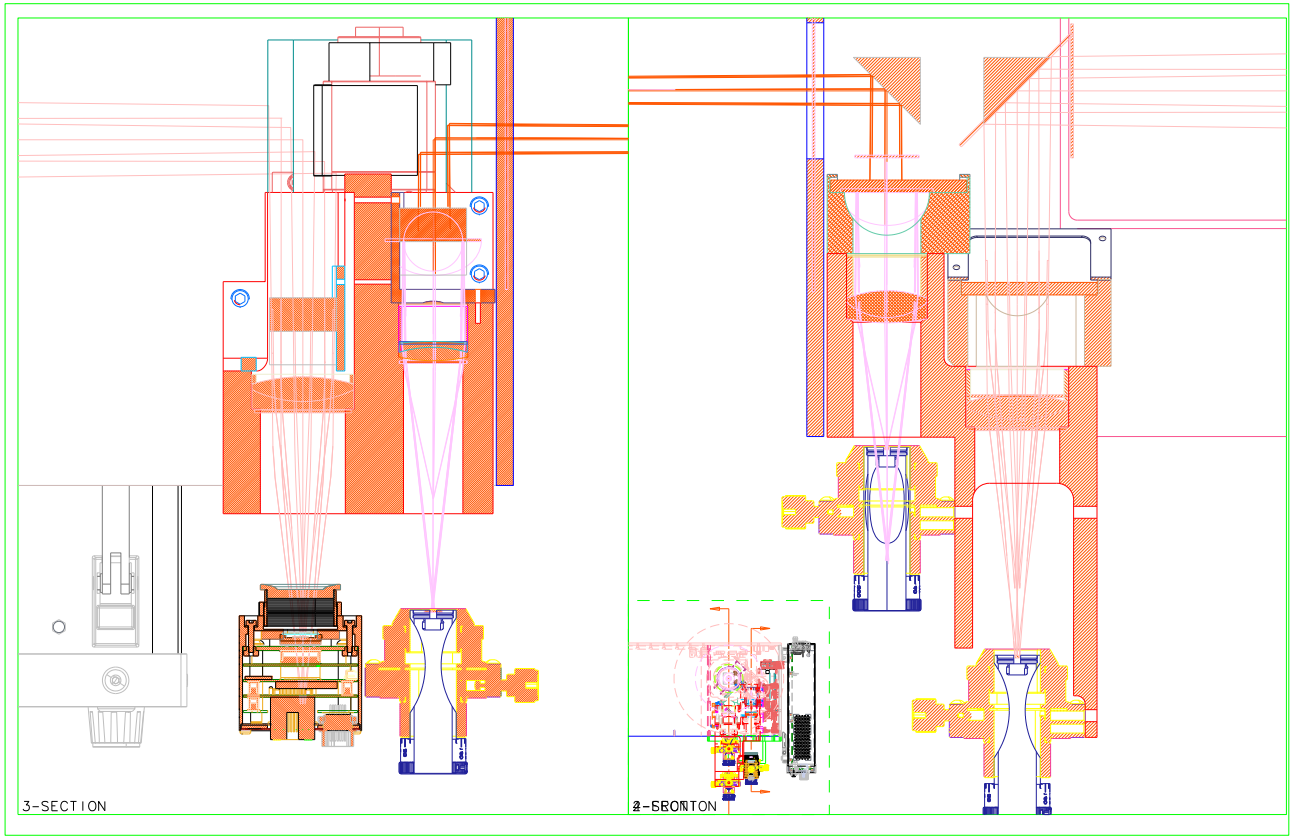}    
      \end{center}
  \caption{Horizontal slices of the mechanical design for the fiber scrambler showing mounts and lens tubes.  The left panel is the slice through the bottom two arms 2 \& 6, showing the fiber acquisition camera and the chuck for the laser diode.  The right panel is the slice through the top two arms 3 \& 4 with the two science fiber chucks at the bottom and the two pickoff mirrors at the top. Optical ray traces are overlaid for reference. For scale, three of the four lenses shown are 1'' in diameter.  \label{fig:f9} }
\end{figure}

\begin{figure}[tb]
  \begin{center}   
    \includegraphics[width=1\textwidth,clip=true,trim=0cm 6cm 0cm 6cm]{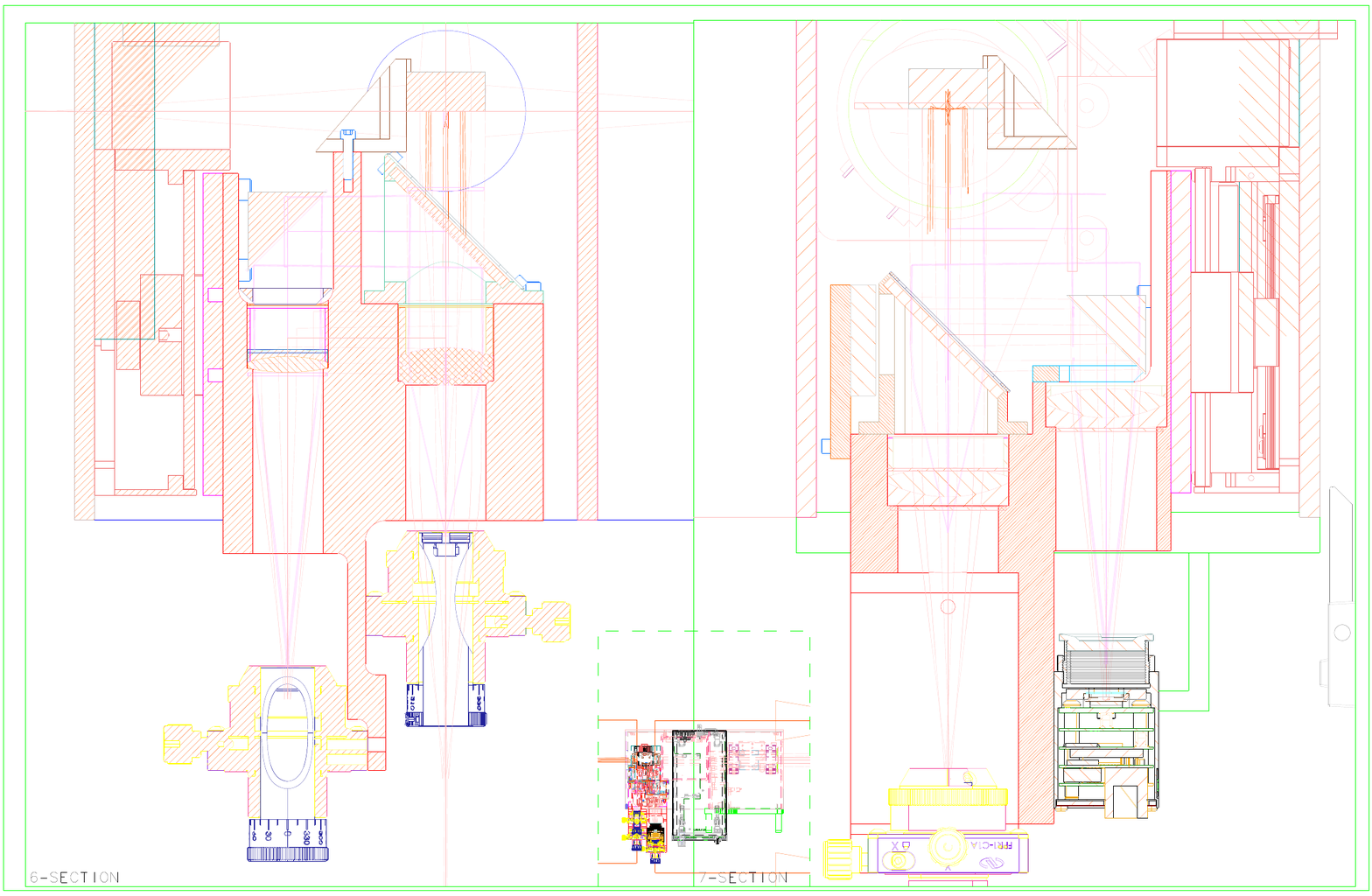}    
      \end{center}
  \caption{Vertical slices of the mechanical design for the fiber scrambler showing mounts and lens tubes, and the stage.  The left panel is the slice through the right two ``output'' arms \#4 \& 6, showing the output science fiber chuck and the fiber chuck for the laser diode.  The right panel is the slice through the left two ''input'' arms \#2 \& 3, showing the input science fiber chuck and the visible alignment camera. Optical ray traces are overlaid for reference.  For scale, three of the four lenses shown are 1'' in diameter. The gas cell is shown as a circle ``head on'' and the fold and cold mirrors are readily distinguishable.  \label{fig:f10} }
\end{figure}

A Newmark Systems, Inc. linear stage (Model MS-2X-E1; 5 kg load) is used to move the fiber scrambler in and out of the beam since it possesses an adequate 2'' travel and fits within the volume constraints.  The stage is mounted to a modified wall of the calibration unit.  The stage is held in place to the wall of the calibration unit via lock screws and slightly over-sized holes in the cover to permit changing the alignment of the stage.   The stage platform was found to be more unstable that desired, and thus we also braced the fiber scrambler against the calibration unit wall perpendicular to the wall to which it is mounted via a metal shim.

The five lenses, two cold mirrors, four fold mirrors, cornercube, visual alignment camera and three fiber chucks to hold the fibers were all mounted to a CNC-machined aluminum block shown in Figure 11.  The main mount was attached to a mounting plate that attached to the linear stage platform.  Lens tubes were hollowed out and custom cylindrical tubes with spring-loaded clamps were machined to hold four of the lenses in place.  Two of the lens tubes are held in place by the custom machined mounts for the cold mirrors for arms 3 \& 4.  The other two lens tubes are held in place by the mounts for the fold mirrors for arms 2 \& 6. All four fold mirrors are glued to their mounts and screwed into the unibody main mount.  The fifth 1'' collimating lens for the cornercube in arm 7 was glued into place after $\sim$5 mm was cut off at a chord to fit the lens into the available space so as not to collide with the cold mirror.  The cornercube is also glued into its mount.  The two cold mirrors are held in place with custom aluminum brackets and mounts.  

The fibers are held in place with three off-the-shelf manual adjustment XYZ$\theta_z$ fiber chucks with four degrees of freedom from Newport Optics (no tip-tilt). A tip degree of freedom is introduced with how the fiber chuck is mounted to the unibody mount via a single screw.  The camera is mounted to the unibody mount with two screws into two machined slots so that the camera focus can be manually adjusted.   The final assembly is shown in Figure 11.

\begin{figure}[tb]
  \begin{center}
    \includegraphics[width=0.55\textwidth]{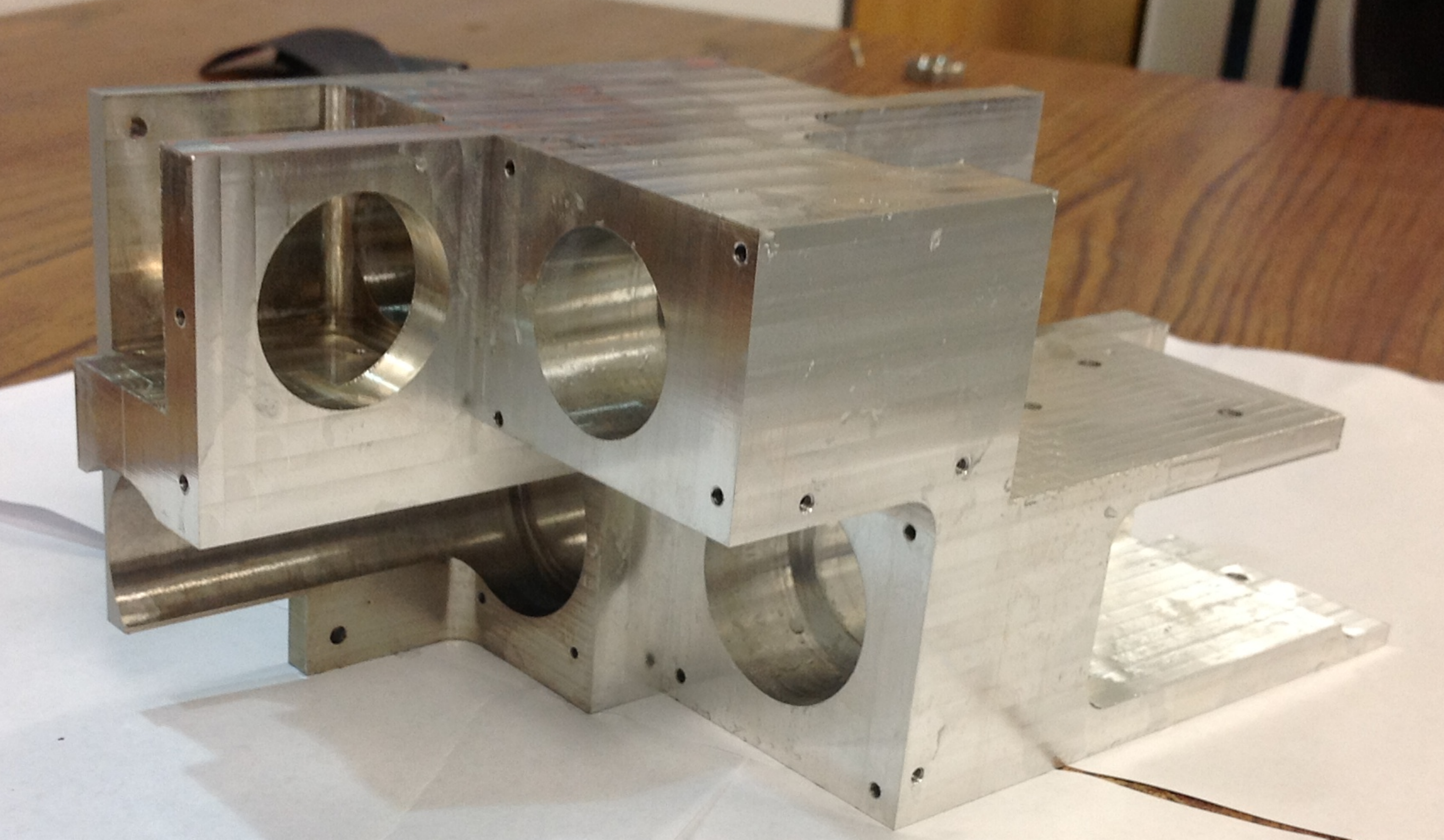}    
    \includegraphics[width=0.35\textwidth]{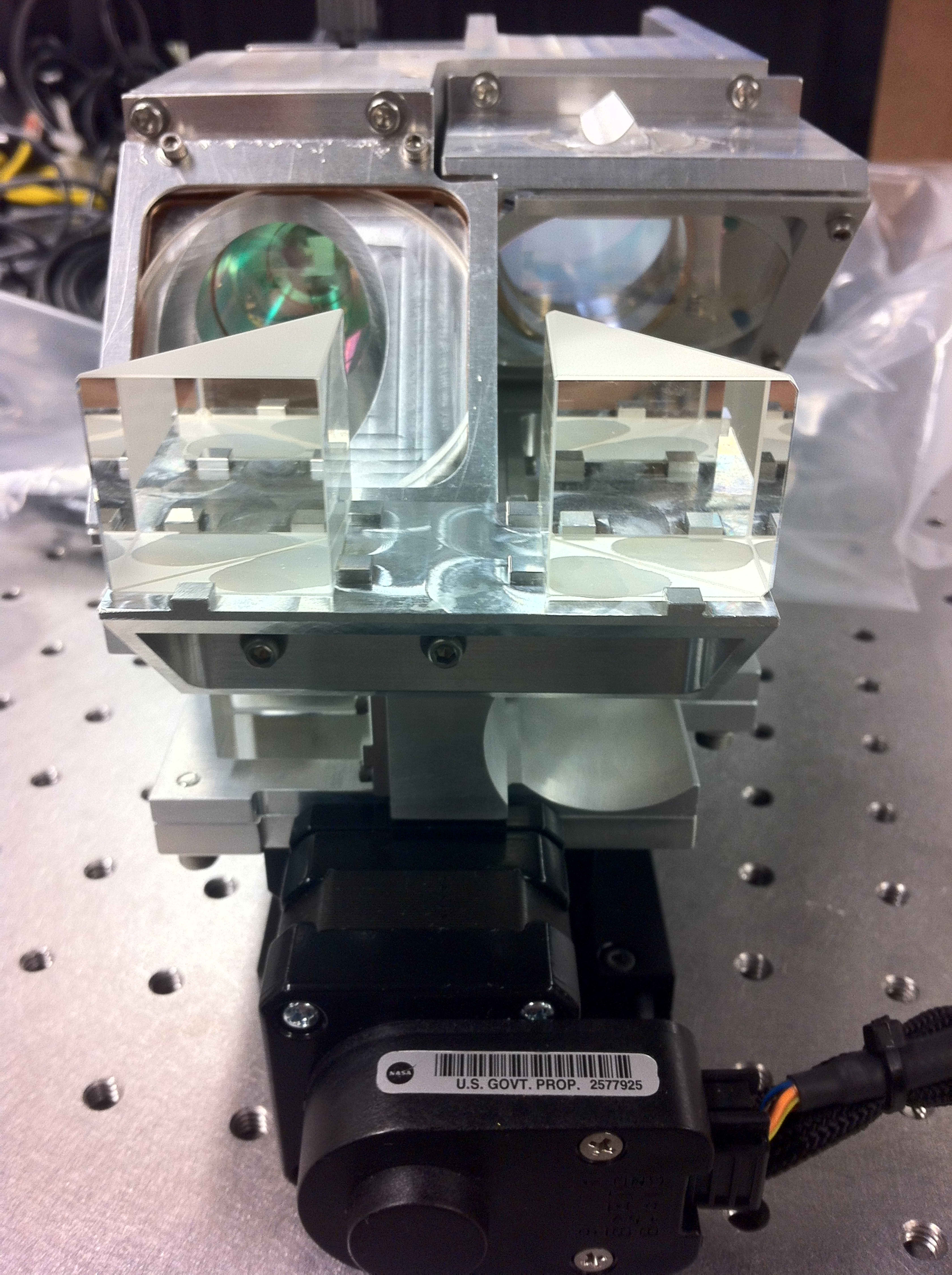}    
      \end{center}
  \caption{Left: Machined unibody mount for all optics and optics holders.  For scale, the lens tubes are 1'' in diameter.  Right: Completed fiber scrambler assembly, viewed from the pickoff mirrors looking towards the two cold mirrors. The cornercube is visible above the input cold mirror, and the stage at the bottom.  \label{fig:f11} }
\end{figure}

Finally, the science output fiber tip displaces visibly on the detector by $\sim$10 microns due to strain in the fiber from cable bends.  Thus, we attached a $\sim$12 inch long metal arm to the unibody mount that extends out from the fiber scrambler to which the science output fiber is clamped at a second and third location in addition to the fiber chuck.  This strain relieves the fiber at the location of the fiber chuck and removes all displacement of the fiber tip on the detector, greatly improving its long-term stability.

\subsection{Fiber Agitation}

One way to mitigate modal noise is to mechanically agitate the fiber \cite{baudrandwalker2001}.  Agitation will cause the spatial and mechanical configuration of the fiber to change, altering the optical path of rays through the fiber, and thus changing the spatial mode illumination distribution at the fiber exit.  If the agitation happens over timescales shorter than a typical exposure time, has a large enough amplitude to cause significant changes in the output illumination, and is sufficiently random, then usually the signal-to-noise can in practice approach the photon-limited regime \cite{lemkecorbettspie2010}.  Shaking the fiber manually by hand seems to be the best method to agitate the fiber that fulfills all of these requirements \cite{2012SPIE.8446E..8JM}.  However, such an agitation mechanism is impractical for actual observations, particularly at high airmasses and over long durations at cold temperatures.  

We find that a good compromise is using ``woofer/tweeter'' setup, where we use a small cell phone vibrator ``tweeter'' motor for high-frequency ($\sim$ 100 Hz), low-amplitude ($\sim$1 mm) oscillations, and a crank arm attached to a planetary gear set from a Japanese toy company to induce low-frequency ($\sim$1 Hz),  high-amplitude ($\sim$ 25 mm)  and low torque oscillations simultaneously.  This electric motor ``woofer'' is mounted to the back of the telescope primary away from the fiber scrambler.  For the 1 m length fibers, use of the ``woofer'' is not practical, and only the ``tweeter'' is used for agitation.

When doing agitation tests in the lab, we also made the following observation: If the agitator is attached at two points along the length of the fiber, then the spatial mode scrambling is more effective compared to when the agitator is attached at a single point along the length of the fiber.  This led us to the conclusion that the optimal agitation method excites a continuous length of fiber simultaneously.  We accomplish this by looping the fiber (at greater than the allowed bending radius of curvature), so that multiple loops of the fiber are attached to the arm of the ``woofer'' agitator motor.  Each loop of fiber reacts like a spring in response to the gravity vector, as it gets longer and narrower and then wider and shorter in an oscillatory manner as the agitator arm turns a full cycle.  

\subsection{Control Electronics and Software}

The remote operation of the fiber scrambler is essential efficient observations, given that CSHELL mounts at the telescope Cassegrain focus.   An electronics box is mounted to the same plate to which the gas cell electronics box is mounted next to the CSHELL instrument\cite{plavchan13}.  The fiber scrambler electronics box contains a RS-232 serial Newmark Systems NSC-1S Single Axis Stepper Motor Controller for the linear stage, the laser, the laser power supply, an ethernet hub, and the laptop power supply.  A commercial laptop is used to issue serial commands to the controller for the linear stage via a USB port, and TCP/IP commands directly to the camera via its built in ethernet-based controller connected to the ethernet hub.  The fiber scrambler control laptop is mounted with the control laptop for the gas cell.  A custom LabVIEW software interface is used to control the camera detector readout, exposure time, etc.  A separate custom LabVIEW software interface is used to control the stage.  The LabVIEW software runs on a Windows operating system with a remote desktop server for remote GUI access from the telescope control room and computers via a remote desktop client.  The laser and fiber agitator are not integrated with the remote operations and control, and require manual activation and de-activation.  We are able to run the agitator for hours each night without interruption or damage to the fibers.

\section{First Light}

The fiber scrambler was commissioned at IRTF on May 7th, 2012.  The first night required manual adjustment to the fiber chucks, camera focus and stage alignment to get starlight to the visual alignment camera and through the fiber to the CSHELL spectrograph.  After that milestone was achieved, relatively routine observations were possible on subsequent nights, with the exceptions of the manual switching of fibers, refocusing the fiber for any wavelength change to the spectrograph, and turning the agitator and laser on and off.  By the end of the first four night observing run, we were able to take series of near-infrared images and spectra of bright standard stars with a variety of our fibers.  Additional runs followed in the summer and fall of 2012.  

Several first light spectra of bright standard stars are shown in Figures 12 \& 13, both in K band at 2.3125 $\mu$m, and in H band at 1.6717 $\mu$m respectively.  All of our radial velocity measurements with the fiber scrambler are made in the H band.  No radial velocity measurements are made at  2.3125 $\mu$m at K band due to three factors impacting the throughput.  The first factor is the transmission loss (Figure 4).  The second factor is slit loss for the 200 $\mu$m fiber.  The third factor is the limited number of modes and poor mode coupling for the 50x100 $\mu$m rectangular and 50 $\mu$m octagonal core fibers.  At 2.3125 $\mu$m, observations of Vega take $\sim$4 minutes to acquire adequate S/N with the 1-m 200 $\mu$m octagonal core fiber. We conclude that any K-band observations with a future fiber-fed spectrograph with a $<$200 $\mu$m core diameter fiber will prove extremely challenging to obtain adequate starlight throughput coupling without the use of adaptive optics.  We arrive at this conclusion regarding the limited number of modes since we did not experience similar mode coupling issues at H band, and since we had relatively good throughput coupling with the 1 m 200 $\mu$m fiber at K band were it not for the slit losses.

\begin{figure}[tb]
  \begin{center}
    \includegraphics[width=0.5\textwidth,clip=true,trim= 0cm 0cm 0cm 4cm]{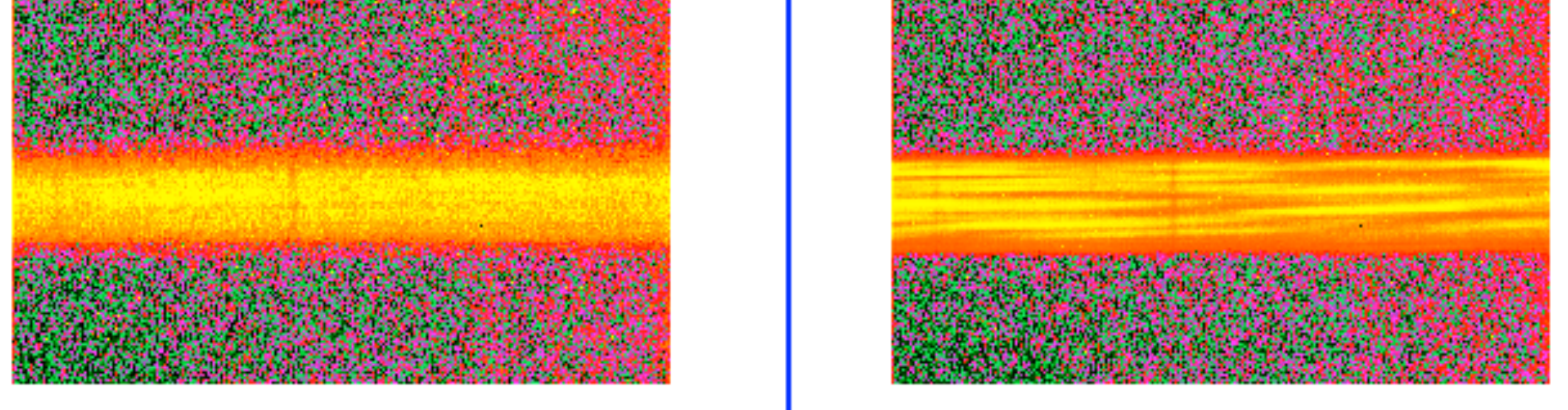}    
    \includegraphics[width=0.5\textwidth]{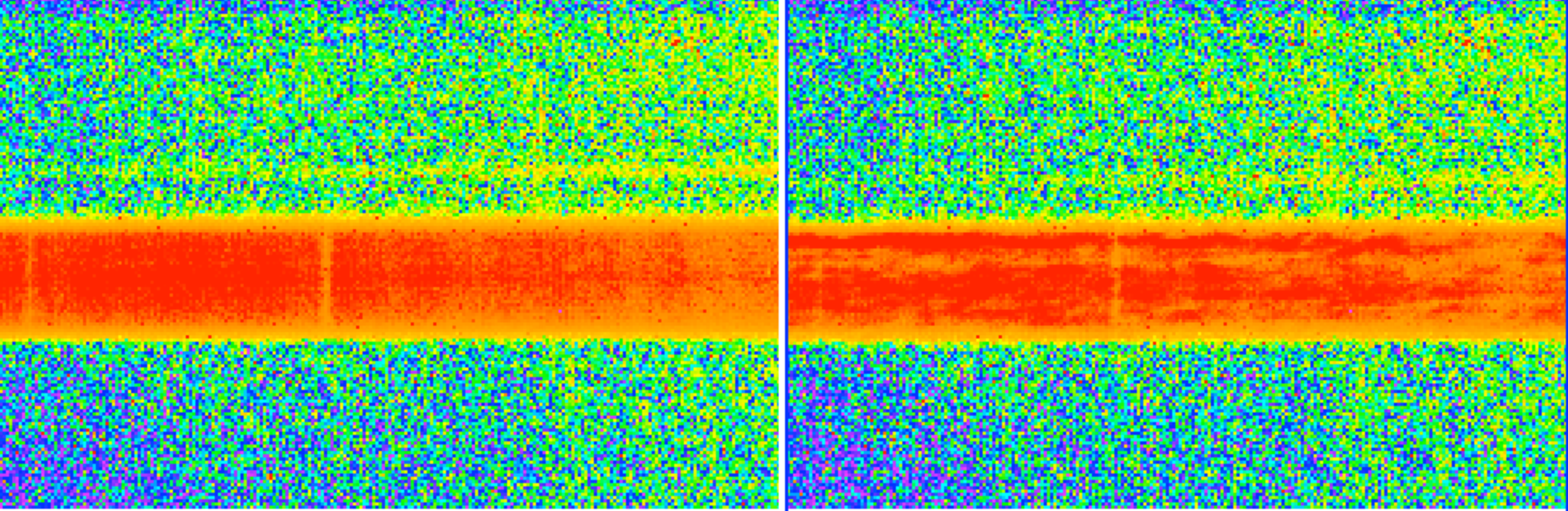}    
      \end{center}
  \caption{Four raw FITS spectra of Vega at 2.3125 $\mu$m, featureless except for two telluric lines (center and left of each spectra).  The vertical direction is the spatial dimension along the slit, and the horizontal dimension is wavelength.  We use 200 $\mu$m octagonal core fibers with a 0.''5 slit that blocks $\sim$90\% of the fiber so that only a central slice of the fiber tip goes through the slit which exacerbates the modal noise present and minimizes the throughput.  First row, left: agitated 1 m fiber; right: non-agitated 1m fiber; second row, left: agitated 10 m fiber; right: non-agitated 10 m fiber.   The benefits of agitation and longer non-circular core fibers are readily apparent in the raw spectra.  The 1 m non-agitated fiber shows strong wavelength-dependent modal noise (changes in the vertical spatial profile of the spectrum from wavelength to wavelength).  The modal noise is less pronounced for the 10 m non-agitated fiber, and for the 1 m agitated fiber.  However, the modal noise is mostly absent for the 10 m agitated fiber (some fringing from the CVF filter is present). \label{fig:f12} }
\end{figure}

\begin{figure}[tb]
  \begin{center}
    \includegraphics[width=1\textwidth,clip=true, trim= 0cm 0cm 16.7cm 0cm]{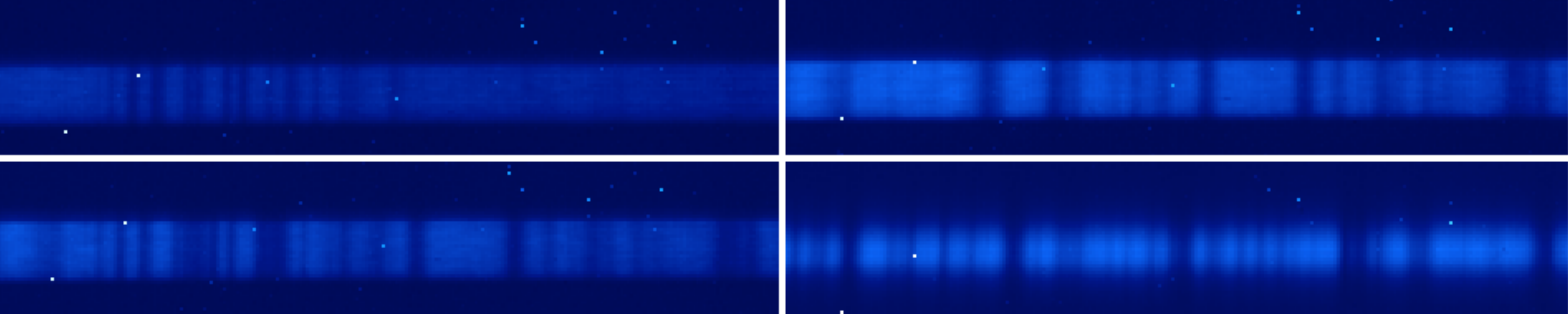}    
      \end{center}
  \caption{Raw FITS spectra of Vega (top left) and SV Peg at 1.6717 $\mu$m.  The vertical direction is the spatial dimension along the slit, and the horizontal dimension is wavelength.   The bottom right panel is a spectrum taken without the fiber scrambler, and the other 3 panels are taken with the fiber scrambler with a 10 m agitated 50x100 $\mu$m rectangular core fiber.  The sharp edges of the fiber are readily apparent compared to the non-fiber spectrum. The illumination stability with the fiber scrambler is not readily apparent by eye.  However, there is a clear lack of modal noise at this shorter wavelength when compared to the K band spectra in Figure 12.  Top left: Vega with the isotopic methane gas cell and fiber scrambler; Top right: SV Peg without the gas cell and with the fiber scrambler; Bottom left:  SV Peg with the fiber scrambler and the gas cell; Bottom right: SV Peg with the gas cell and without the fiber scrambler.  \label{fig:f13} }
\end{figure}

\section{Preliminary Results}

Several hundred high S/R spectra of the K=-0.4 mag supergiant M7 star SV Peg were obtained with and without the fiber scrambler at 1.6717 $\mu$m, with S/N and exposure times of 95$\pm$12, 30--45 s and 61$\pm$13, 1.75 s for the fiber and non-fiber observations respectively.   Future publication will present the analysis for the remaining fiber scrambler observations, and the performance of the data pipeline\cite{bottom13}.  An overview of the data pipeline, including the optimal spectral extraction and forward modeling to extract radial velocities is presented in Plavchan et al. (2013).  The pipeline is a significantly evolved version from Anglada-Escude et al. 2012, and was developed over the past two years to overcome a number of significant data challenges identified since the preliminary analysis of first light data.   As mentioned in ${\S}$2, the spectrograph CVF filter produces fringing in our spectra at the few percent level in flux.  While we include a model for the fringing in our radial velocity extraction pipeline, the fringing nevertheless introduces a long-term systematic noise source.  

Figure 14 shows preliminary radial velocity measurements for the red supergiant SV Peg taken at 1.6717$\mu$m with and without the fiber scrambler.  The observations taken with the fiber scrambler utilize an agitated 50x100 $\mu$m rectangular core 10 m fiber.  The fiber observations bracket the non-fiber observations in time and airmass.  A long-term linear trend has been removed from both radial velocity curves, and this trend is correlated with an incomplete fringing correction in the stellar template in our analysis to date.   We can co-add these measurements to probe the short-term noise floor with and without the fiber scrambler as a function of S/N.  For the fiber data, a precision of 27 m/s is measured for a S/N=95$\pm$12.  This precision decreases to $\sim$12 m/s for S/N$>$350.     For the non-fiber data, a precision of 126 m/s is measured for a S/N=61$\pm$13.  This precision decreases to $\sim$25 m/s for S/N$>$350.   From these measurements, we can conclude that the fiber scrambler benefits the obtainable RV precision for a given S/N on short time-scales, by stabilizing the illumination of the spectrograph and consequently making the output spectra and their variations easier to model.  The fiber scrambler obtains significantly better precision for S/N$\sim$100.  Part of this is likely due to the short 1.75 s integrations without the fiber scrambler, where the PSF and consequently the LSF variations are more pronounced relative to the fiber data taken with longer exposure times.

\begin{figure}[tb]
  \begin{center}
   \includegraphics[width=0.45\textwidth]{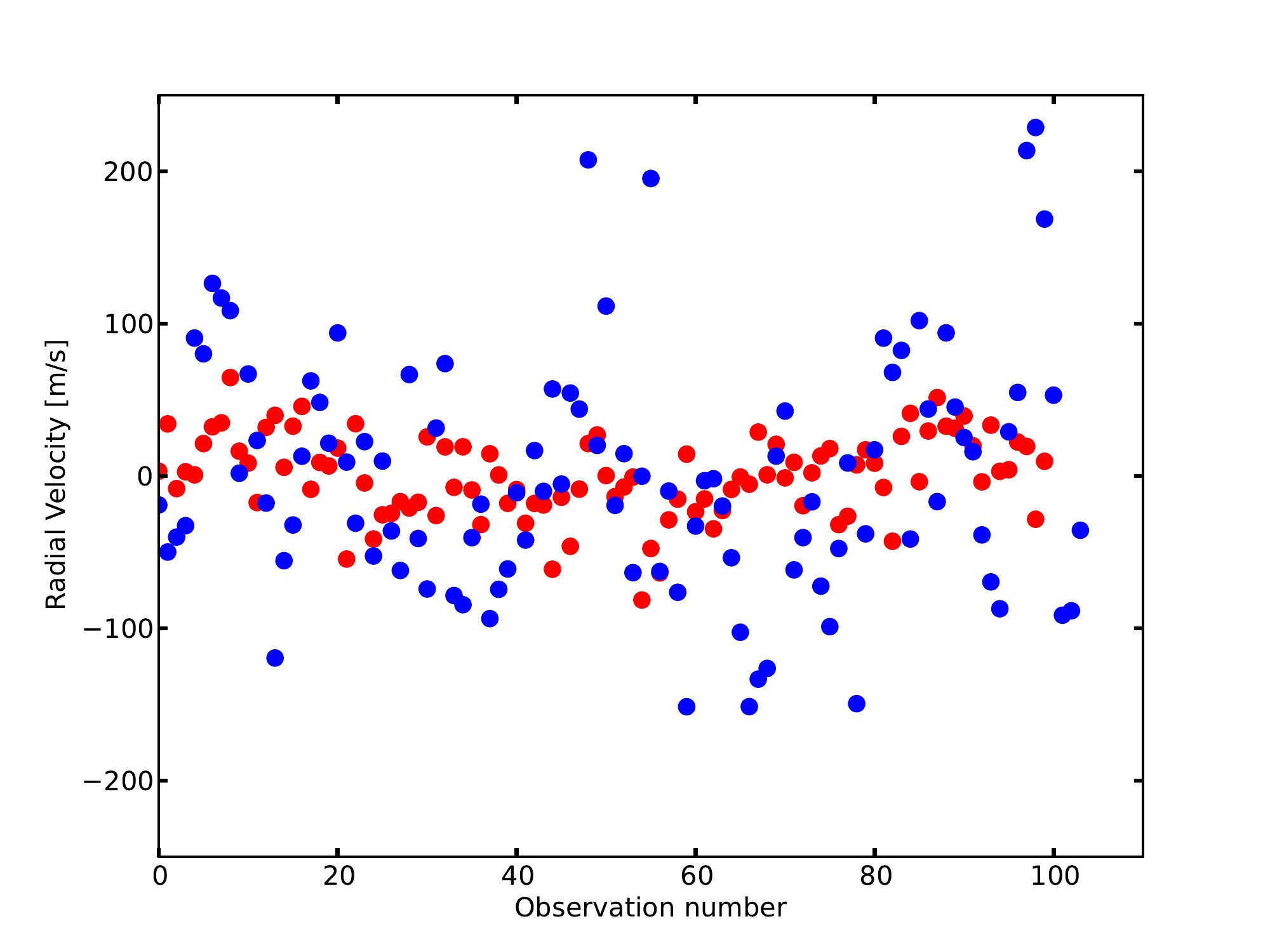}    
    \includegraphics[width=0.45\textwidth]{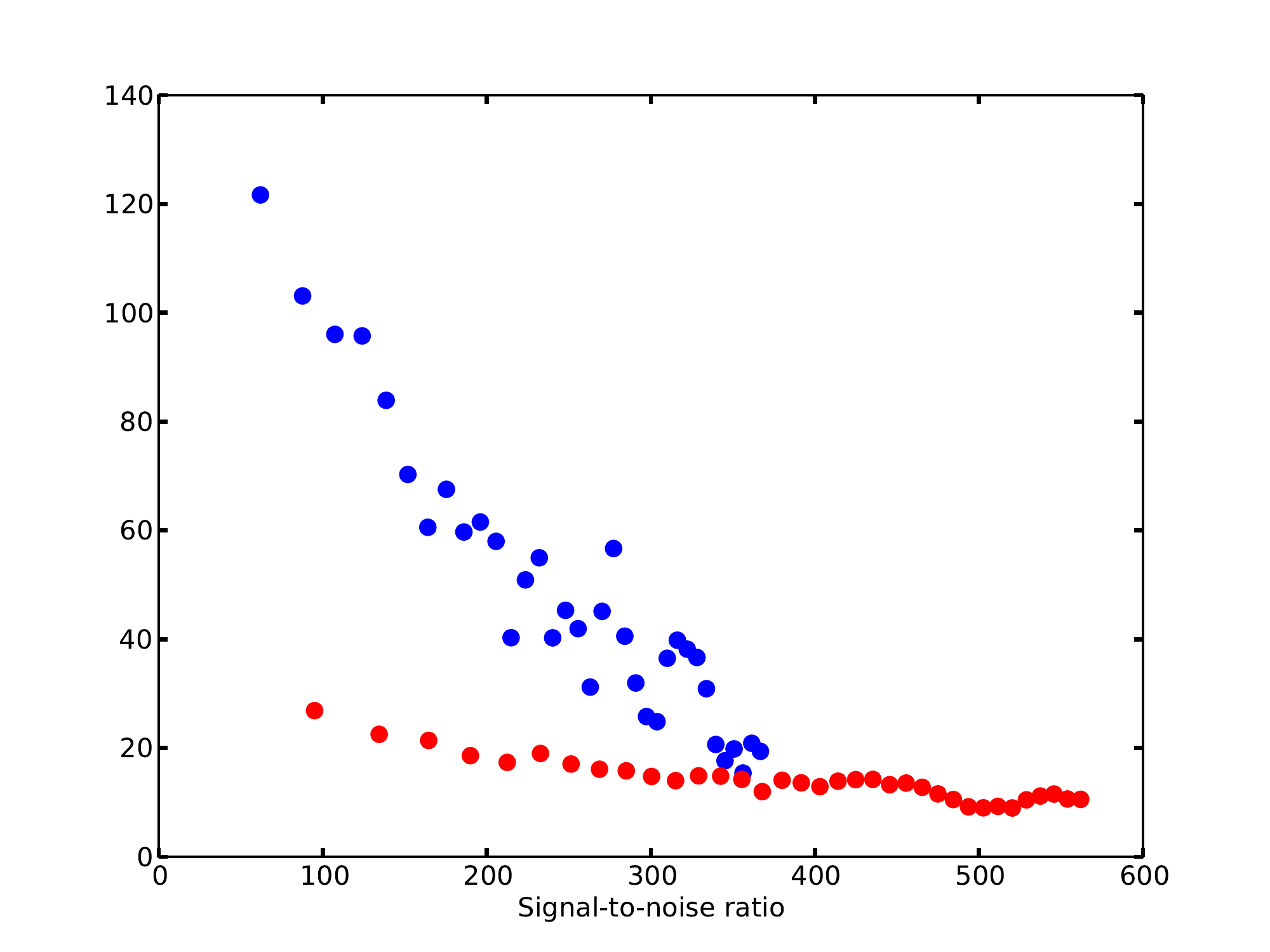}    
    \includegraphics[width=0.45\textwidth]{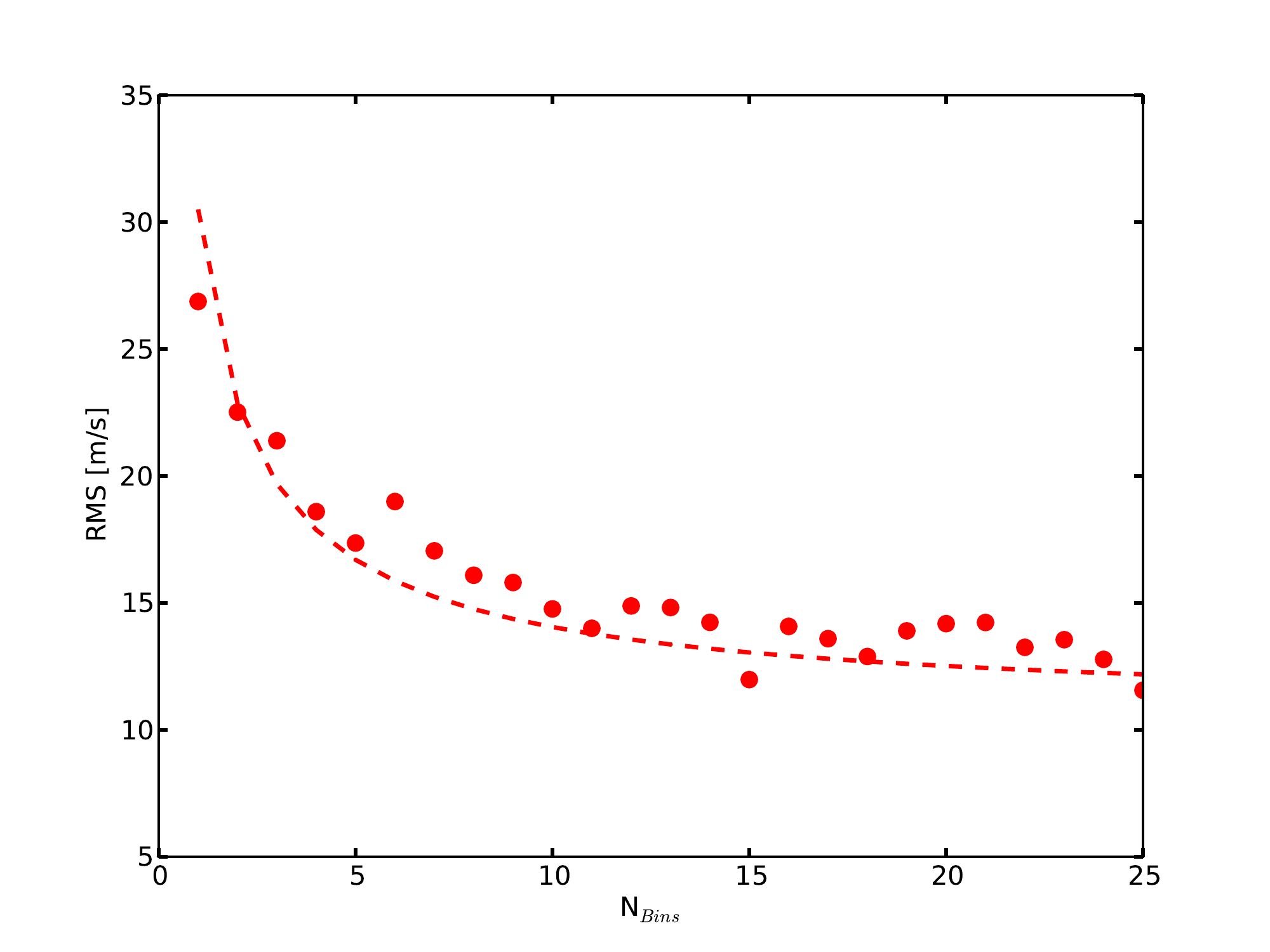}    
    \includegraphics[width=0.45\textwidth]{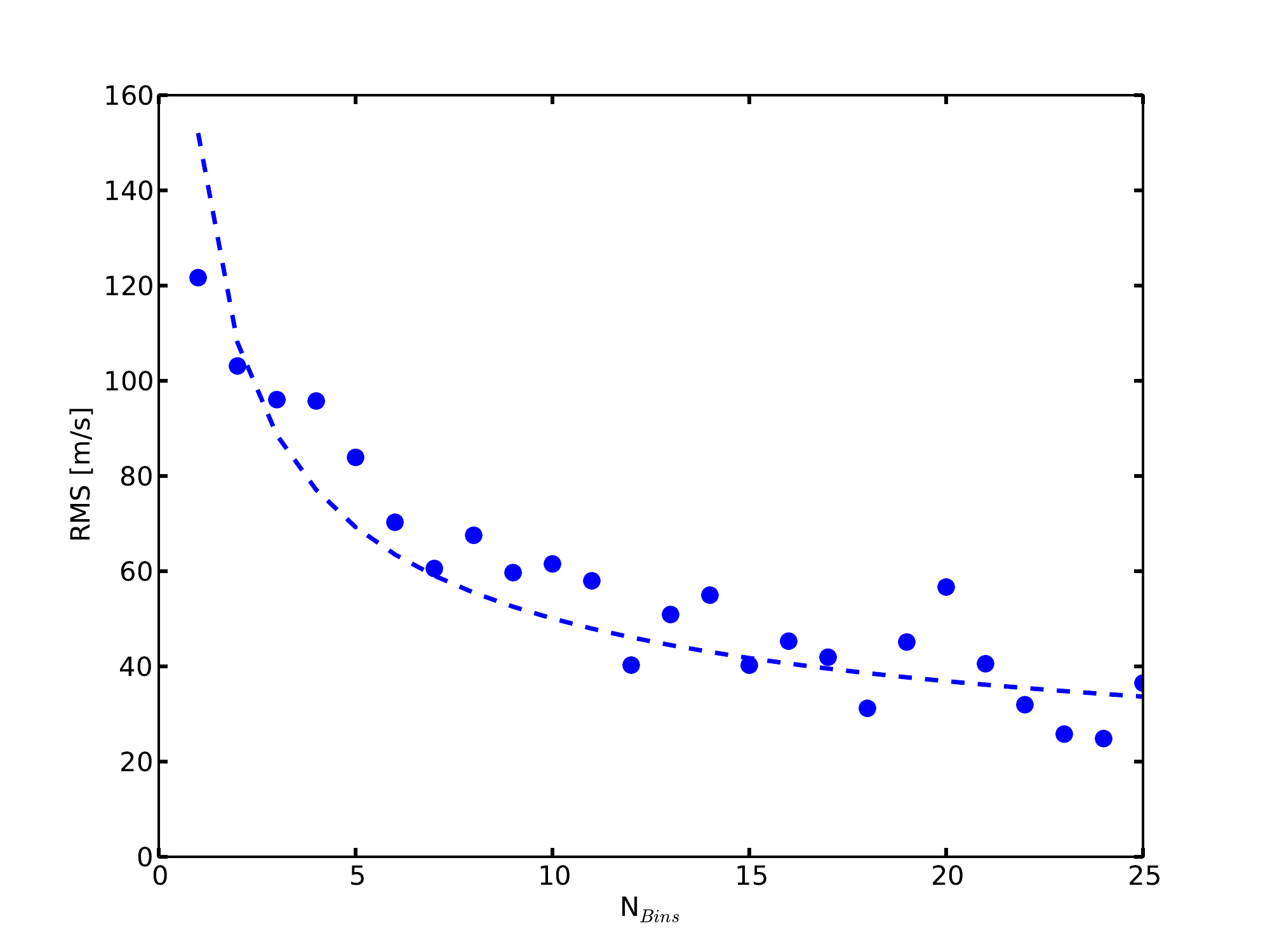}    
      \end{center}
      
\caption{Top Left: Relative radial velocity measurements of the super giant star SV Peg (K = -0.4 mag) obtained during the commissioning runs.  The horizontal axis corresponds to the observation number, and the vertical axis the radial velocity.  The fiber scrambler data are shown in red, and the non-fiber data are shown in blue.  The non-fiber data has been scaled by the ratio of the relative S/N of the fiber and non-fiber data (divided by 95/61) to effectively match the S/N per observation of the fiber-scrambler data.   Linear drifts in the radial velocity from both the fiber and non-fiber data have been divided out, and are attributed to an incomplete fringing correction in the stellar template.  Top Right: We coadd every $M$ radial velocity measurements ($M$=1,2,3, etc.), and we recompute the radial velocity precision.  The horizontal axis is the effective S/N, which goes as $\sqrt{M}$, and the vertical axis is the corresponding radial velocity precision.   The fiber scrambler data are shown in red, and the non-fiber data in blue.  The co-added radial velocity precisions decrease to $\sim$12 m/s with the fiber scrambler and $\sim$25 m/s without.   These plots demonstrate that the fiber radial velocity measurements are better than the non-fiber radial velocity measurements for the same effective S/N.  Bottom Left: The same radial velocity noise curve data shown in the top right panel for the fiber data, with a white noise and noise floor model fit to the data with amplitudes of 28.54 and 10.77 m/s respectively shown as the red dashed line.  The horizontal axis is the number of binned measurements for a given radial velocity precision, and is proportional to the (S/N)$^2$.  Bottom Right: The corresponding radial velocity noise curve data for the non-fiber data.  The corresponding white noise and noise floor model fit to the data is shown as a blue dashed line with amplitudes of 151 and 14.7 m/s respectively\cite{muirhead}. \label{fig:f14}}
\end{figure}

\section{Conclusions and Future Work}

We have built and commissioned the first near-infrared fiver scrambler with non-circular core fibers to work with the CSHELL spectrograph at IRTF.  We have demonstrated a short time-scale RV noise-floor for suitably bright targets at H-band of $\sim$12 m/s.   We have demonstrated that fiber scramblers can be extended from the visible to the near-infrared, and hold promise for input illumination stabilization on planned future near-infrared spectrographs with more modern detectors and cross-dispersion for larger spectral grasp, including iSHELL on IRTF, an upgraded NIRSPEC on Keck, iGRINS, the Habitable Zone Planet Finder, SPIRou and others.

\acknowledgments     
 
Peter Plavchan would like acknowledge Wes Traub and Stephen Unwin for seed funding provided by the JPL Center for Exoplanet Science and NASA Exoplanet Science Institute, as well as JPL Research and Technology Development grant in FY13.  G. Anglada-Escud\'e would like to acknowledge the Carnegie Postdoctoral Fellowship Program and the support provided by the NASA Astrobiology Institute grant NNA09DA81A.  Part of the research at the Jet Propulsion Laboratory (JPL) and California Institute of Technology was performed under contracts with National Aeronautics and Space Administration.  We would like to thank Steve Osterman (U. of Colorado) for their valuable discussions. We also thank John Rayner, Morgan Bonnet, George Koenig, Kars Bergknut and Alan Tokunaga from IfA/Hawaii for their support during the fiber scrambler design reviews, integration and commissioning.  


\bibliography{report}   
\bibliographystyle{spiebib}   

\end{document}